\documentclass[11pt]{article}

\usepackage[english]{babel}

\usepackage[letterpaper,top=2cm,bottom=2cm,left=3cm,right=3cm,marginparwidth=1.75cm]{geometry}

\usepackage{amsmath, amsthm}
\usepackage{amsfonts}
\usepackage{graphicx}
\usepackage[colorlinks=true, allcolors=blue]{hyperref}
\usepackage{graphicx}
\usepackage[normalem]{ulem}

\usepackage{amssymb}
\usepackage{latexsym}
\usepackage{amsmath,amstext,amsthm,amssymb,bbm,amsbsy}
\usepackage{mathtools}
\usepackage{tabularx}
\usepackage{subcaption}
\usepackage{dsfont}
\usepackage{xcolor}
\usepackage{comment}

\newtheorem{proposition}{Proposition}[section]
\newtheorem{remark}[proposition]{Remark}
\newtheorem{assumption}[proposition]{Assumption}
\newtheorem{example}[proposition]{Example}

\title{Solving Optimal Execution Problems via \\ In-Context Operator Networks}

\author{Tingwei Meng\footnote{University of California Los Angeles, Department of Mathematics, Los Angeles, CA 90095, USA.}
\hspace{3ex}
Moritz Voss\footnotemark[1] \footnote{Corresponding author.}
\hspace{3ex}
Nils Detering\footnote{Heinrich Heine University D\"usseldorf, Mathematisches Institut, Universitätsstra{\ss}e 1, 40225 D\"usseldorf, Germany.}  \vspace{.5em} \\
Giulio Farolfi\footnotemark[1]
\hspace{3ex}
Stanley Osher\footnotemark[1]
\hspace{3ex}
Georg Menz\footnotemark[1]
}

\begin{document}

\maketitle

\begin{abstract}
We propose a novel transformer-based neural network architecture (ICON-OCnet) for solving optimal order execution problems in the presence of unknown price impact. Our architecture facilitates data-driven in-context operator learning for the incurred price impact by merging offline pre-training with online few-shot prompting inference. First, the operator learning component (ICON) learns the prevailing price impact environment from only a few executed trade and price impact trajectories (time series data) provided as context. Second, we employ ICON as a surrogate operator to train a neural network policy (OCnet) for the optimal order execution strategy for the price impact regime inferred from the in-context examples. We study the efficiency of our approach for linear propagator models with path-dependent transient price impact and explicitly known optimal execution strategies. In this model class, price impact persists and decays over time according to some propagator kernel. We illustrate that ICON is capable of accurately inferring the underlying price impact model from the data prompts, even for propagator kernels not seen in the training data. Moreover, we demonstrate that ICON-OCnet correctly retrieves the exact optimal order execution strategy for the model generating the in-context examples. Our introduced methodology is very general, offering a new approach to solving path-dependent optimal stochastic control problems sample-based with unknown state dynamics.   
\end{abstract}

\section{Introduction}

Devising optimal order execution strategies for buying or selling large volumes of shares of a stock on a centralized exchange is a major concern for large institutional investors and hence became a well-studied problem in quantitative finance in the past two decades. The aim is to split up a large meta order into smaller child orders which are executed over some time horizon to mitigate adverse price impact incurred by large trades. We refer to the monographs~\cite{CarteaJaimungalPenalva:15, Gueant:16, BouchaudEtAl:18, Webster:23} for a comprehensive overview of this topic. Typically, these problems are addressed as optimal stochastic control problems by first formulating a price impact model that describes how trades affect the execution price, and then solving for an optimal execution strategy tailored to this model.

In contrast, our approach is concerned with the limitations of postulating a specific parametric model in real-world trading applications, where market conditions are non-stationary and vary over time, and stylized price impact parameters capturing the interaction with the market when executing trades are difficult to estimate in a reliable and persistent manner. To tackle this challenge, we propose a novel sample-based, in-context operator learning framework, termed ICON-OCnet. Specifically, we adapt the In-Context Operator Networks (ICON) learning methodology proposed in~\cite{YangLiuMengOsher:23} to (i) learn the \emph{unknown} price impact regime prevailing in a {\em current} market environment from only a few recently executed sample trades, and then (ii) devise the associated optimal order execution strategy, a neural network optimal control policy (called OCnet), for the detected price impact environment. 

We examine our general approach within the broad class of linear propagator models proposed by~\cite{BouchaudEtAl:04, BouchaudEtAl:09, Gatheral:10}, where we can effectively benchmark our methodology against the ground truth optimal execution strategies recently obtained in~\cite{JaberNeuman:25}. Propagator models constitute a versatile class of transient price impact models defined by a price impact kernel (propagator), which captures, in reduced form, the interplay between price moves and current and past trades as empirically observed when executing market orders in limit order books. Note, however, that the method developed in this paper does not depend on this particular model setup and can be applied to any price impact framework.  

Our proposed ICON-OCnet method is based on an \emph{in-context learning paradigm}, in other words, (transfer-)learning from prompted time series example data serving as context. To this end, we leverage a transformer-based architecture, which is very well suited for sequential data. The idea is to infer the price impact operator, which maps the trading rate as a function of time to the incurred price impact path, using only a few observed sample trajectories. This allows the transformer to adjust to new or unseen price impact models without requiring full retraining of the network. First, we train the ICON model offline on synthetically generated trade and price impact paths from a wide range of propagator model specifications. The pretrained ICON model is then initialized with only a few examples of trades and corresponding realized price impact to infer the underlying price impact operator in a few-shot online learning manner from the prompted context. In a second step, we feed the initialized ICON model as a surrogate operator into the optimal order execution problem and train a neural network policy via a policy gradient method akin to~\cite{HanE:16} to learn the optimal execution strategy for the price impact operator inferred by ICON from the provided context. We validate the performance of our ICON-OCnet approach through various numerical experiments, demonstrating its ability to (i) detect the true price impact operator even when trained on data from a different propagator model class, and (ii) correctly recover the corresponding optimal order execution strategy from the propagator model generating the in-context examples. In particular, the ICON surrogate operator demonstrates sufficient precision and robustness to be effectively utilized in an iterative stochastic gradient descent-type optimization algorithm to learn the corresponding optimal policy. 

Finally, the broader purpose of this paper is also to demonstrate a \emph{proof of concept} for our general methodology in a complex and non-trivial setting: solving a path-dependent optimal control problem with unknown state dynamics, inferred in a data-efficient and robust manner from a limited number of in-context examples by leveraging the few-shot and transfer learning capacities of transformer networks. 

Related to our work, in the sense of addressing a similar problem but model-specific for linear propagator price impact models, are the studies~\cite{NeumanZhang:23} and~\cite{NeumanStockingerZhang:23}: \cite{NeumanZhang:23} introduces a statistical online learning approach for linear propagator models, alternating between exploration and exploitation phases, and achieving sublinear regret with high probability;  \cite{NeumanStockingerZhang:23} proposes an offline learning framework to estimate the price impact kernel while accounting for uncertainty in the estimator and derives the asymptotic optimality of strategies in terms of execution cost. In contrast, we suggest in-context, few-shot learning, which efficiently combines offline and online learning, enhancing model flexibility while limiting the need for costly online training. Moreover, our approach is fully model-agnostic and only assumes that the price impact is a function (operator) of the trading trajectory. In particular, our setup is not confined to linear propagator models, and we do not perform any model parameter estimation or classical kernel estimation as in~\cite{NeumanStockingerZhang:23, NeumanZhang:23}. Instead, we leverage the capabilities of the transformer-based neural network architecture in ICON to detect commonalities in the prompted in-context examples of trade and price impact trajectories with the sample paths encountered during offline training to perform a reliable and stable price impact prediction of executed trades that can be successfully employed in solving a downstream optimal execution problem. To the best of our knowledge, the only work that uses transformer models for trade execution is~\cite{kim2023adaptive}, which develops an adaptive dual-level reinforcement learning framework based on a combined transformer and Long-Short-Term Memory (LSTM) architecture to track the Volume-Weighted Average Price (VWAP). There has recently been also a lot of interest in the general learning of solution operators for ordinary and partial differential equations with varying network architectures. Among many others, we mention approaches based on Deep Operator Networks (DeepONet) in~\cite{Luetal:21}, graph kernels in~\cite{anandkumar-kernel}, Fourier Neural Networks in~\cite{anandkumar-fourier2}, and structure-informed operator learning in~\cite{BenthDeteringGalimberti:24,benth2024}.

The rest of the paper is organized as follows: Section~\ref{Intro:propagator:model} introduces the general optimal order execution problem. Our ICON-OCnet approach is described in Section~\ref{sec:ICON-OCnet}. Numerical results are presented in Section~\ref{sec:numerical}. Conclusion and outlook are summarized in Section~\ref{sec:conclusion}. 

\section{Optimal order execution problem} \label{Intro:propagator:model}

In this section, we summarize the classical optimal order execution problem in the presence of price impact within a unifying and fairly generic modeling framework that is commonly used in the literature, encompassing various introduced price impact models. We refer to~\cite{CarteaJaimungalPenalva:15, BouchaudEtAl:18, Webster:23} for general reference.

\subsection{Model setup} \label{sec:modelsetup}

Let $T>0$ denote a finite deterministic time horizon and fix a filtered probability space $(\Omega,\mathcal{F},\mathbb{F}:=(\mathcal{F}_t)_{0 \leq t \leq T}, \mathbb{P})$ satisfying the usual conditions. We consider a trader with an initial inventory $x \in \mathbb{R}$ in a risky asset (e.g., number of shares held in a stock) who wishes to liquidate her position by time $T$, which represents, for instance, the end of the trading day. The trader's inventory, i.e., the number of shares held at any time $t \in [0,T]$, is modeled by the process $X = (X_t)_{0 \leq t \leq T}$ with dynamics
\begin{equation} \label{def:X}
    X_t = x - \int_0^t u_s \, ds \qquad (0 \leq t \leq T),
\end{equation}
where $u = (u_t)_{0 \leq t \leq T}$ denotes her selling rate chosen from the set of admissible strategies
\begin{equation} \label{eq:admissibleset}
    \mathcal{U} := \left\{ u : \mathbb{F}\text{-progressively measurable s.t.} \int_0^T \mathbb{E}[u^2_t] \, dt < \infty \right\}. 
\end{equation}
We assume that the trader's trading activity causes price impact in the sense that her trades at time $t \in [0,T]$ are filled at the execution price 
\begin{equation} \label{eq:ExecutionPrice}
    P_t = S_t - Y_t \qquad (0 \leq t \leq T).
\end{equation}
Here, $S=(S_t)_{0 \leq t \leq T}$ denotes a square integrable special semimartingale and represents the unaffected price process, i.e., the price process that would have prevailed without the trader's trading, modeling price changes caused by other traders and the arrival of new information. In contrast, the process $Y=(Y_t)_{0 \leq t \leq T}$ models the \emph{price impact} incurred by the trader due to her trading strategy $u$ such that $Y_t$ depends on her current and past order flow $(u_s)_{0 \leq s \leq t}$ for all $t \in [0,T]$. We assume $Y_0 = 0$ so that $P_0 = S_0$ in~\eqref{eq:ExecutionPrice}.\footnote{It is also conceivable to allow for a general $Y_0 = y \in \mathbb{R}$. In this case, the starting value $y$ models an initial dislocation of $P_0$ from the fundamental value $S_0$ at the opening of the continuous trading session on $[0,T]$.} 

More precisely, we make the following general assumption.
\begin{assumption} \label{ass:standing}
    We assume that there exists a determinstic, non-anticipative price impact operator $\boldsymbol{I}_\theta: L^2([0,T],\mathbb{R}) \rightarrow L^2([0,T],\mathbb{R})$ satisfying
    \begin{equation} 
        (\boldsymbol{I}_{\theta}(u))_t = (\boldsymbol{I}_{\theta}( u \vert_{[0,t]}))_t \qquad (0 \leq t \leq T)
    \end{equation}
    for all $u \in \mathcal{U}$, such that the trader's price impact process $Y$ in~\eqref{eq:ExecutionPrice} is given by 
    \begin{equation} \label{eq:operatorI}
        Y_t = (\boldsymbol{I}_{\theta}(u))_t \qquad (0 \leq t \leq T).
    \end{equation}
\end{assumption}

Henceforth, we will also write the trader's execution price $P$ in~\eqref{eq:ExecutionPrice} as
\begin{equation} 
    P_t = S_t - (\boldsymbol{I}_{\theta}(u))_t \qquad (0 \leq t \leq T).
\end{equation}
In particular, we interpret the operator $\boldsymbol{I}_{\theta}(\cdot)$ as encoding the \emph{price impact environment} in which the trader executes her trades, and we use the subscript $\theta$ for convenience to emphasize the dependence on a certain price impact \emph{regime}. Note that we do not use the term model here to emphasize that $\boldsymbol{I}_{\theta}(\cdot)$ does not necessarily need to represent a parametric price impact model with some parameter vector $\theta$. For now, we deliberately introduce $\boldsymbol{I}_{\theta}$ without any further specification to highlight the versatility of our approach. Concrete examples will be considered later in our numerical study in Section~\ref{sec:numerical}; cf.~Example~\ref{ex:linearpropagators}.   

\subsection{Optimization problem} \label{subsec:optimization}

The trader's goal is to find an optimal order scheduling strategy $u \in \mathcal{U}$ that maximizes her objective functional 
\begin{align} \label{eq:objective}
J(u) : = & \, \mathbb{E}\left[ \int_0^T P_t u_t \, dt - \varepsilon \int_0^T u_t^2 \, dt - \phi \int_0^T X_t^2 \, dt + X_T S_T - \varrho X_T^2 \right] \nonumber \\
= & \, \mathbb{E}\left[ \int_0^T \big( S_t - (\boldsymbol{I}_{\theta}(u))_t \big) u_t \, dt - \varepsilon \int_0^T u_t^2 \, dt - \phi \int_0^T X_t^2 \, dt + X_T S_T - \varrho X_T^2 \right]
\end{align}
for some constants $\varepsilon > 0$, $\phi \geq 0$ and $\varrho \geq 0$; cf., e.g., \cite{LehalleNeuman:18, NeumannVoss:22, JaberNeuman:25} and the references therein.  The first integral in~\eqref{eq:objective} represents the trader's gains from her selling strategy $u$. The second integral with constant $\varepsilon > 0$ describes in reduced form all additional \emph{instantaneous costs}, which are not directly linked to persistent price impact but incurred, for instance, by the bid-ask spread. The constants $\phi \geq 0$ and $\varrho \geq 0$ implement, respectively, a penalty on the running and terminal inventory. In particular, a large value for $\varrho$ virtually enforces the desired liquidation constraint that $X_T$ will be very close to zero. Lastly, the term $X_T S_T$ represents the final asset position's value in terms of the unaffected fundamental price $S_T$. Observe that $J(u) < \infty $ for any admissible $u \in \mathcal{U}$. 

It is also very common and helpful to rewrite the performance functional in~\eqref{eq:objective} as 
\begin{equation} \label{eq:objective2}
    J(u) = \mathbb{E}\left[ \int_0^T \big( \alpha_t - (\boldsymbol{I}_{\theta}(u))_t \big) u_t \, dt - \varepsilon \int_0^T u_t^2 \, dt - \phi \int_0^T X_t^2 \, dt - \varrho X_T^2 \right] + x \mathbb{E}[S_T]
\end{equation}
to emphasize the tradeoff of the execution problem between exploiting the asset's short term \emph{intraday alpha signal}
\begin{equation} \label{eq:alpha}
    \alpha_t := \mathbb{E}[S_t - S_T \vert \mathcal{F}_t] \qquad (0 \leq t \leq T)
\end{equation}
and the incurred price impact $(\boldsymbol{I}_{\theta}(u))_t$ from liquidation; see~\cite[Chapter 2]{Webster:23}. This signal predicts the expected future returns of the unaffected price $S$ during the trading period $[0,T]$. For the optimal execution problem, it serves as a sufficient statistic of the unaffected price process~$S$. Hence, the predictor $\alpha = (\alpha_t)_{0 \leq t \leq T}$ is usually considered as the relevant additional input variable to the execution problem which is modeled by the controller (trader), rather than $S$ itself; see also the discussion in~\cite[Section 3.2]{Hey:25}.

To summarize, the aim of the trader is to solve the optimal stochastic control problem 
\begin{equation} \label{def:optproblem}
    J(u) \rightarrow \max_{u \in \mathcal{U}}.
\end{equation}

Observe that the objective function in~\eqref{eq:objective} or, equivalently, in~\eqref{eq:objective2} is fairly generic and does not hinge on a specific price impact model. Indeed, the operator $\boldsymbol{I}_{\theta}(\cdot)$ therein can be any price impact operator. Generally speaking, our goal in the remainder of this paper is to develop a methodology that numerically computes the optimal strategy in~\eqref{def:optproblem} when the price impact operator $\boldsymbol{I}_{\theta}(\cdot)$ in~\eqref{eq:objective2} is an \emph{unknown} ``black-box'' operator that is inferred only from a few trading examples.

\begin{remark} \label{rem:literature}
We emphasize that in the literature on optimal order execution, it is very common to set up a price impact model by postulating a decomposition of the execution price~$P$ as in~\eqref{eq:ExecutionPrice}, under the assumption that the unaffected price process $S$ is known and, as a consequence, the impact process $Y$ is observable. Of course, on real financial markets, only the affected execution price $P$ is observed, but not $S$ or $Y$. As we will explain below, our methodology can be employed when only the observed execution price $P$ is available. 
\end{remark}

\begin{remark} \label{rem:determinstic}
Note that, due to the representation of the performance functional in~\eqref{eq:objective2}, the optimal liquidation strategy depends on the intraday alpha signal $(\alpha_t)_{0 \leq t \leq T}$ in~\eqref{eq:alpha}, which is very sensible. The process $\alpha$ is the only source of randomness in the objective that needs to be modeled by the trader. Moreover, if the unaffected price process $S$ in~\eqref{eq:ExecutionPrice} is a martingale, we have $\alpha \equiv 0$ in~\eqref{eq:alpha}, and the optimization problem in~\eqref{def:optproblem} reduces to a deterministic control problem. In this case, the trader does not have any directional view on the unaffected price process $S$, and the optimal order scheduling strategy is a deterministic function of time $t \in [0,T]$, depending only on the price impact operator $\boldsymbol{I}_{\theta}$, as well as the hyperparameters $\varepsilon, \phi, \varrho$, and initial inventory $x$.  
\end{remark}

\section{Price impact operator learning and neural network solver} \label{sec:ICON-OCnet}

\begin{figure}[htbp]
    \centering
    \begin{subfigure}{.8\textwidth}
        \centering \includegraphics[width=\textwidth]{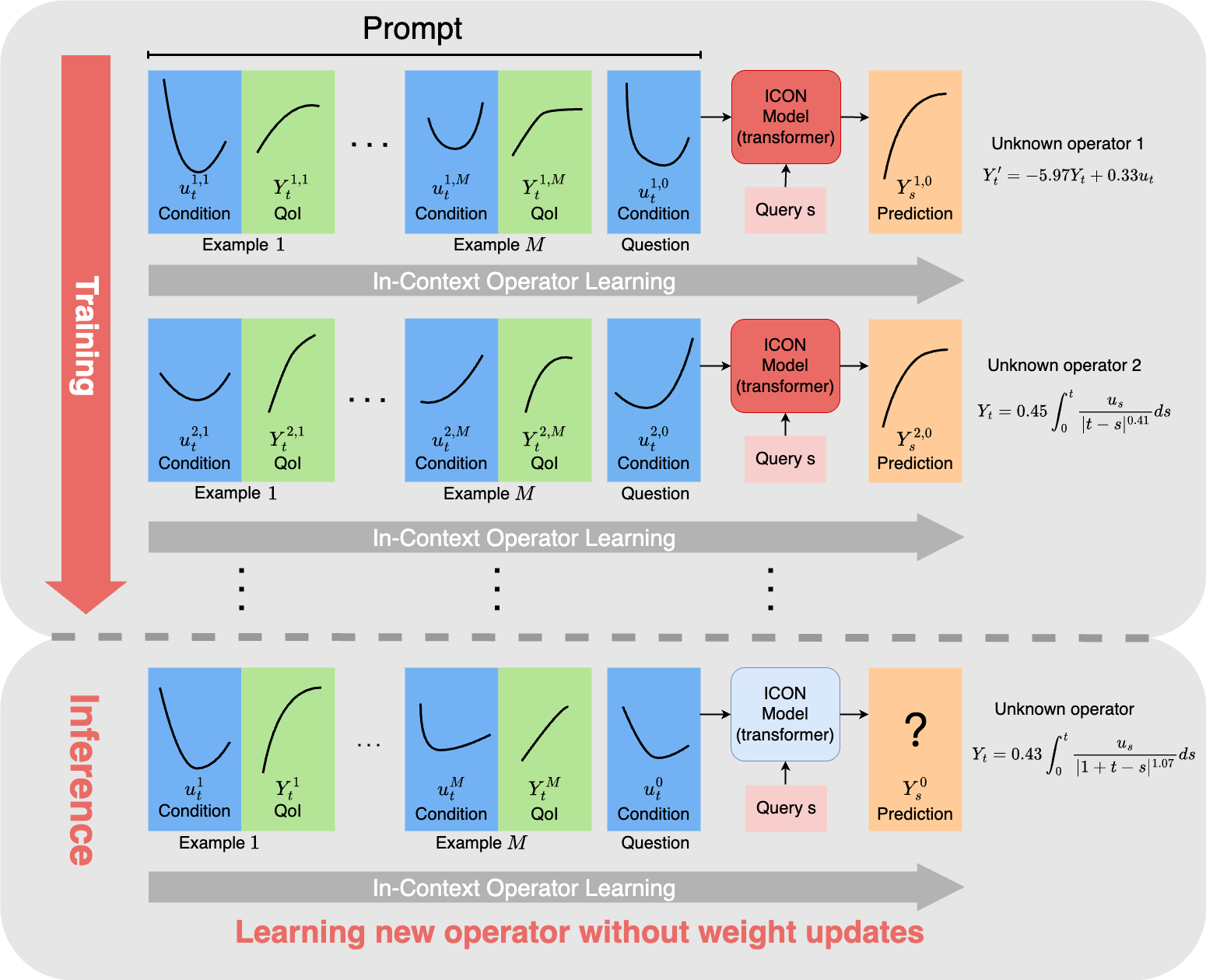}
        \caption{Step 1: ICON training}
    \end{subfigure}
    
    \vspace{3em}
    
    \begin{subfigure}{.8\textwidth}
        \centering \includegraphics[width=\textwidth]{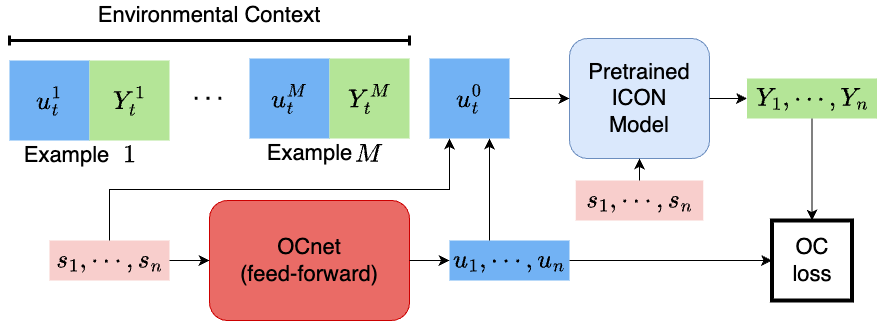}
        \caption{Step 2: OCnet training}
    \end{subfigure}
    \caption{Illustration of the ICON-OCnet structure. In both steps, the red rounded rectangle represents the training of a neural network, while the blue rounded rectangle indicates a pre-trained neural network with frozen parameters.}
    \label{fig:illustration_icon-oc}
\end{figure}

In this section, we describe the in-context price impact learning task and downstream optimization algorithm used to find the optimal liquidation strategy in~\eqref{def:optproblem} with unknown price impact operator~$\boldsymbol{I}_{\theta}$. An illustration of the method, which we coin ICON-OCnet, is shown in Figure~\ref{fig:illustration_icon-oc}. It consists of two steps summarized as follows:

\begin{enumerate}
\item Train ICON offline in a data-driven way on a range of price impact models. Then, use the pre-trained ICON model in a few-shot online learning manner to infer the operator $\boldsymbol{I}_{\tilde{\theta}}$ mapping $u$ onto $Y$ in~\eqref{eq:operatorI} based on only a few examples stemming from a specific model $\tilde{\theta}$, possibly not seen during offline training. Regarding the time series data used as \emph{in-context examples}, there are two possible approaches: one can either use discretized trajectories of pairs $(u,Y)$ or directly prompt discretized realizations of $(u,P)$. For the former, since $Y$ is not directly observable in practice, it is assumed that the price impact $Y$ is filtered out from the (observed) execution price $P$ in~\eqref{eq:ExecutionPrice} in a pre-processing step.\footnote{We refer to the related discussion of fitting price impact models in~\cite[Chapter 7]{Webster:23} and the references therein.} 
\item Train a neural network policy (OCnet) via a policy gradient method to approximate the optimal execution strategy in~\eqref{def:optproblem} based on the ICON surrogate operator $\boldsymbol{\hat{I}}_{\tilde{\theta}}$ obtained in step~1. In particular, the performance measure for training OCnet is a discretization of the objective functional in~\eqref{eq:objective2} with $(\boldsymbol{I}_{\tilde{\theta}}(u))_t$ replaced by the in-context prediction $(\boldsymbol{\hat{I}}_{\tilde\theta}(u))_t$ of ICON.
\end{enumerate}

To wit, by prompting only a few samples of time series data of selling rates and corresponding incurred price impact trajectories, our method derives the optimal execution strategy for the unknown price impact environment that generates the examples.

\subsection{ICON training and few-shot learning} \label{subsec:ICON}

Concerning step 1, inspired by~\cite{YangLiuMengOsher:23}, we use In-Context Operator Networks (ICON), a novel transformer-based neural network architecture designed to learn the operator $\boldsymbol{I}_{\theta}: L^2([0,T],\mathbb{R}) \rightarrow L^2([0,T],\mathbb{R})$ defined in~\eqref{eq:operatorI} that maps the trading (selling) rate process $u$ to the price impact process $Y = \boldsymbol{I}_\theta(u)$. The network is pre-trained in an offline learning step and can then, in a second online learning step, detect the price impact operator from a few provided examples. Unlike traditional neural network methods that require re-training or fine-tuning for each new problem, the so-called few-shot learning or prompting technique used here requires only a few example trades and their respective price impact in order to recognize the underlying relationship between $u$ and $Y$. In particular, it can deal with new, unseen price impact models without actually retraining the network weights. This allows the network to handle a wide range of price impact models by leveraging commonalities shared between different models. From a practical perspective, this methodology has several advantages. Specifically, the true relationship $u \mapsto Y$ is typically unknown. Even after a decision on a specific model has been made, there might be uncertainty about the parameters of the model. Moreover, the model parameters might not be constant over time. With the in-context few-shot learning approach, a trading desk tasked with liquidating a large position can, in principle, base its search for an optimal execution strategy in a data-driven way on {\em experience} from recent trades of the same asset. In this way, the network architecture automatically detects, from a universe of possible price impact models, a model that best fits the most recently {\em incurred} price impact as provided by the few examples or test trades as context.  

Using a transformer architecture also offers multiple benefits for our few-shot prompting task in the context of optimal liquidation. First, its attention mechanism allows the neural network to efficiently process inputs of varying lengths, making it well suited for handling different types of examples (data prompts) and key-value pairs (i.e., time instances $s_j$ and observations $(u_{s_j},Y_{s_j})$, respectively $(u_{s_j},P_{s_j})$) without necessitating structural changes. More precisely, the transformer accepts inputs up to a maximum sequence length, with shorter sequences padded and masked accordingly. From a practical point of view, this flexibility is very helpful in learning the true price impact operator. Indeed, the trades and the corresponding observed price impact or execution prices that are used as examples are usually not uniform in length and are sampled at different, possibly non-equidistant and fairly irregular time steps. Second, the dependence of the output on the input of transformers can be quite flexible, and hard constraints can be added through the design of specific masks. For instance, our relationship between $u$ and $Y$ is non-anticipative; hence, we use a modified causal mask to add this constraint. Moreover, and most importantly for our purposes, the multi-head attention mechanism of the transformer is very well suited for \emph{sequential time series data} and enables capturing the relationships between different parts of the input. This facilitates the learning of the price impact by focusing on the most relevant aspects of the examples. As a consequence, transformers can deal with complex interdependencies, making them a very natural choice for a typically \emph{non-Markovian} price impact environment, which relates price moves intertemporally to current and past trades. Lastly, transformers also support parallel processing, rendering it computationally efficient to generate predictions across multiple query points simultaneously, which is vital for fast inference in few-shot settings. We also refer to the discussion in~\cite{YangLiuMengOsher:23}. 

\subsubsection{ICON training with in-context pairs \texorpdfstring{$(u,Y)$}{(u,Y)}} \label{subsec:training:uY:based}

We first explain the training procedure and use-case for an idealized situation where the trading-incurred price impact $Y$ is perfectly known and available as context for offline training and online inference. For the sake of clarity, this will also be the setup in our proof-of-concept experiment with synthetic time series data in Section~\ref{sec:numerical} below. The corresponding in-context learning procedure and few-shot prompting inference of the ICON model are illustrated in Figure~\ref{fig:illustration_icon-oc}~(a). 

First, each row above the dashed line represents one data point that is used in training. The data in row $i$ is produced by the same price impact model $\theta_i$ and consists of $M$ ``condition'' and ``quantity of interest (QoI)'' pairs of discretized trajectories $(u^{i,1},Y^{i,1}), \ldots, (u^{i,M}, Y^{i,M})$, observed on possibly irregular time grids $t^{i}_1,\ldots,t^{i}_{l^i}$, serving as the context, followed by the ``question'' (or ``question condition'') $u^{i,0}$ specified on some time grid $s_1, \ldots, s_k$, the associated ``query'' $(s_1,\ldots,s_k)$, and the corresponding ``prediction'' $(Y^{i,0}_{s_1},\ldots,Y^{i,0}_{s_k})$. The query provides ICON with the grid information on which the prediction is evaluated. Note that the discretization for the question condition $(s_1,\ldots,s_k)$ can differ from the discretization of the example pairs $(t^i_1,\ldots,t^i_{l^i})$, $i \in \{1,\ldots,M\}$, which may even vary across the $M$ examples themselves. Based on this data, ICON is trained in an offline manner on different models. In Figure~\ref{fig:illustration_icon-oc}~(a), this is illustrated by the different price impact operators of different linear propagator models, which we use in our numerical study in Section~\ref{sec:numerical}. We use a suitable self-attention mechanism via masking to ensure that $Y$ is learned and predicted in a non-anticipative way with respect to $u$ (in the provided examples and in the prediction for the question condition); see also the discussion below. 

Next, at the inference stage illustrated below the dashed line in Figure~\ref{fig:illustration_icon-oc}~(a), the trained ICON network is presented with $M$ new time series sample data (data prompts) of $(u,Y)$ pairs, stemming from a possibly unseen model~$\tilde\theta$, discretized on some time grid. Based on these examples, the pre-trained ICON network infers the underlying price impact operator $\boldsymbol{I}_{\tilde\theta}$ and provides the output prediction for an arbitrary question condition $u^0$ at an arbitrary query $(s_1, \ldots, s_n)$. In other words, it predicts the realized price impact $Y^0 = \boldsymbol{I}_{\tilde\theta}(u^0)$ incurred by a strategy $u^0$ on the time grid $(s_1, \dots, s_n)$ for a {\em possibly unknown model} $\tilde\theta$ that generated the sample time series data as context. Moreover, predictions of the price impact $\boldsymbol{I}_{\tilde\theta}(u)$ for any strategy $u \in \mathcal{U}$ on a varying time grid can then be obtained by providing as context the same $M$ example data pairs from the underlying model $\tilde{\theta}$ in every inference step. In this way, we obtain an ICON \emph{surrogate operator} $\boldsymbol{\hat{I}}_{\tilde\theta}$ for the true operator $\boldsymbol{I}_{\tilde{\theta}}$ that we can feed into an optimization algorithm to compute the corresponding optimal order execution strategy in~\eqref{def:optproblem}. The subscript $\tilde\theta$ in the notation for the surrogate operator $\boldsymbol{\hat{I}}_{\tilde\theta}$ emphasizes that the provided context (examples) originates from the same model $\tilde\theta$. We stress, however, that ICON does not actually infer the underlying model $\tilde\theta$ and its parameters. Instead, it internally processes an encoder-based representation of the model underlying the examples, which is based on similarities with the models generating the examples in the offline training step. In particular, the model $\tilde\theta$ could be a completely new, unseen model that is even outside the class of models encountered in training, as long as this family of models is rich enough to embed important characteristics of $\tilde\theta$.

\begin{figure}[htbp]
\centering
\includegraphics[width=.8\textwidth]{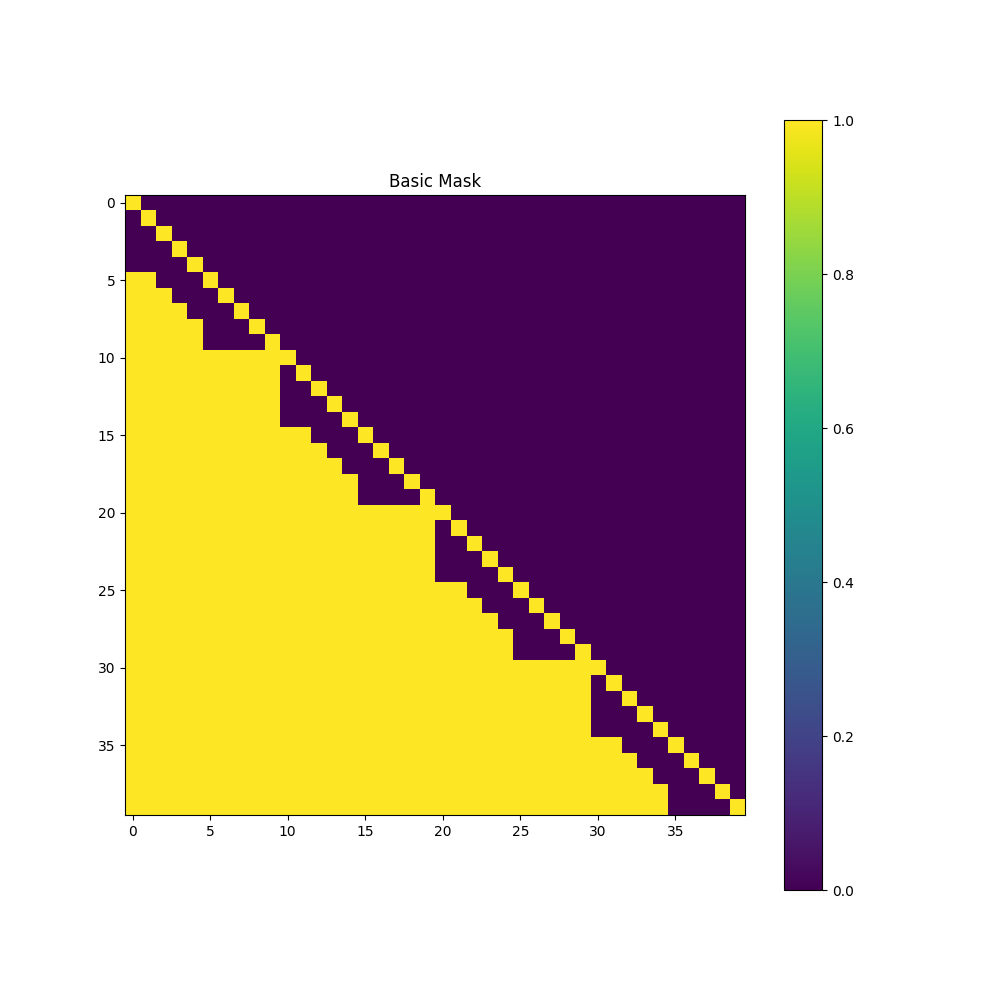}
\caption{Representation of the attention mask used in ICON during the inference step in a simplified setup with only $M=3$ example pairs $(u^m, Y^m)_{m=1,\ldots,3}$ and a fixed time grid $s_0, s_1, \ldots, s_4$ with $s_0 = 0$. More precisely, in all three examples, the condition consists of $(Y^m_{s_0}, u^m_{s_0}, u^m_{s_1}, u^m_{s_2}, u^m_{s_3})$ and the quantity of interest is given by $(Y^m_{s_0}, Y^m_{s_1}, Y^m_{s_2}, Y^m_{s_3}, Y^m_{s_4})$. Note that we effectively include the starting value $Y^m_0$ as the first value in the condition (for the sake of generality, even though we assume $Y_0=0$ throughout) and in the QoI. The desired prediction is then $Y^0_{s_0}, Y^0_{s_1}, Y^0_{s_2}, Y^0_{s_3}, Y^0_{s_4}$ with query $(s_0, s_1, s_2, s_3, s_4)$ for a given question condition $(Y^0_{s_0}, u^0_{s_0}, u^0_{s_1}, u^0_{s_2}, u^0_{s_3})$. The indices of the input sequence are arranged on the vertical axis from $0$ to $39$, and from the horizontal axis one can read off the indices it ``self-attends'' to (yellow regions). More precisely, the input is arranged as one sequence, starting with the three examples $(Y^1_{s_0}, u^1_{s_0}, \ldots, u^1_{s_3}, Y^1_{s_0}, \ldots, Y^1_{s_4}, \ldots, Y^3_{s_0}, u^3_{s_0}, \ldots, u^3_{s_3}, Y^3_{s_0}, \ldots, Y^3_{s_4})$ (indices 0-29), followed by the question condition $Y^0_{s_0},u^0_{s_0},\ldots,u^0_{s_3}$ (indices 30-34) and query times~$s_0,\ldots,s_4$ (indices 35-39). From the graph we see that the trained model self-attends the value $Y^m_{s_i}$ in the $m$-th example only to past values $u^m_{s_j}$ with $j \leq i$ and the starting value $Y^m_{s_0}$, without access to future information (neither in $u^m$ nor $Y^m$); as well as to all previous example pair values $(Y^n_{s_0}, u^n_{s_0}, \ldots, u^n_{s_3}, Y^n_{s_0}, \ldots, Y^n_{s_4})$, $n < m$, on the entire grid, serving as additional context for the $m$-th example. The same logic applies for the actual prediction of the value $Y^0_{s_i}$ for the query input $s_i$, $i \in \{0, \ldots, 4\}$, in the last five rows (input indices 35-39), where only the values $Y^0_{s_0}$, $u^0_{s_j}$, $j \leq i$, in the input are attended (from the indices 30-34), as well as all previous example pair values $(u^m_{s_l}, Y^m_{s_l})_{m=1,\ldots,3}$ (indices 0-29) serving as the context.
}
\label{fig:mask_train_test}
\end{figure}

Our ICON neural network architecture corresponds to the one used in~\cite{YANG2025107455} for learning solution operators of partial differential equations. To implement the non-anticipative structure within the transformer, we construct a causal (self-)attention mask that enforces the correct temporal dependency between the control $u$ and its response $Y$, in the provided examples serving as context as well as in the actual question condition. That is, the mask ensures that, when learning/predicting a value $Y_{s_i}$ at some time instant $s_i$, the model can only self-attend to the past control inputs $u_{s_j}$ at instances $j \leq i$, thereby preserving the causal relation between the trajectories $u$ and $Y$, and preventing any information leakage from the future. Moreover, multiple example pairs are arranged within one sequence so that the model simultaneously learns from several input-output relations. This sequence design allows earlier examples to also serve as context for later examples, and hence enables the transformer to learn consistent causal dependencies across varying temporal grids. We present the attention mask in a simplified setup for the online inference step (i.e., step 1 in Figure~\ref{fig:illustration_icon-oc} (a) below the dashed line) in Figure~\ref{fig:mask_train_test}, where the yellow regions indicate allowed self-attention and the dark regions indicate masked connections. The block lower-triangular structure reflects the causal, non-anticipative nature of the inference process, as well as the fact that earlier examples also serve as context for later examples as mentioned above; see also the caption of Figure~\ref{fig:mask_train_test} for more details. The mask we use for training ICON (i.e., for step 1 in Figure~\ref{fig:illustration_icon-oc} (a) above the dashed line) is more involved and not presented here. More details on the training are provided in Sections~\ref{subsec:data} and~\ref{subsec:ICONnumerics}.

\subsubsection{ICON training with in-context pairs \texorpdfstring{$(u,P)$}{(u,P)}} \label{subsec:training:uP:based}

We now outline the changes to the offline and online training procedures when time series data of pairs $(u,P)$ of trading strategies $u$ and corresponding realized execution price trajectories $P$ in~\eqref{eq:ExecutionPrice} are to be used as the actual in-context examples. In this case, ICON can, in principle, be trained to filter out the price impact from the context. Note, however, that in contrast to~$Y$, the realizations of $P$ are noisy due to the additional presence of the unaffected price process $S$. 

More precisely, as long as the offline training step is model-based, using synthetic time series data of the execution price $P$ in~\eqref{eq:ExecutionPrice} with known trading-incurred price impact $Y$ as in Section~\ref{subsec:training:uY:based} above (and a known model for $S$), one can generate a training set that consists of (noisy) realizations $(u^{i,1},P^{i,1}), \ldots, (u^{i,M}, P^{i,M})$ serving as the $M$ context examples, followed by the question condition $u^{i,0}$, a query $(s_1,\ldots,s_k)$, and the prediction $(Y^{i,0}_{s_1}, \ldots, Y^{i,0}_{s_k})$ of the true price impact. Here, $Y^{i,0}$ is consistent with the model generating the in-context execution prices $(P^{i,1}, \ldots, P^{i,M})$. Based on this data, ICON can be trained offline on a wide range of different price impact models as described above and illustrated in Figure~\ref{fig:illustration_icon-oc} (a), except that the in-context example pairs $(u^{i,1},Y^{i,1}), \ldots, (u^{i,M}, Y^{i,M})$ are now replaced by realizations $(u^{i,1},P^{i,1}), \ldots, (u^{i,M}, P^{i,M})$. For the online few-shot inference step, the trained ICON network is then presented with $M$ new example trajectories $(u^1, P^1), \ldots, (u^M, P^M)$ and predicts the realized price impact $Y^0 = \boldsymbol{I}_{\tilde\theta}(u^0)$ of an arbitrary prompted strategy $u^0$ on some query $(s_1,\ldots,s_n)$, based on the detected impact model $\tilde\theta$ underlying the in-context examples, by leveraging commonalities with the learned models from the offline training.\footnote{In practice, proprietary trading data could serve as prompted examples, each consisting of a meta-order execution strategy that was executed over some time period via a sequence of smaller child orders.}

Generally speaking, the offline training and online inference procedure of ICON can be seen as being conditional on the underlying model in the prompted examples. When offline training is based on in-context examples $(u^{i,1}, Y^{i,1}), \ldots, (u^{i,M}, Y^{i,M})$ as outlined in Section \ref{subsec:training:uY:based} above, this conditional training and inference is a classical operator learning problem for the deterministic operator $\boldsymbol{I}_{\tilde\theta} : L^2([0,T]) \rightarrow L^2([0,T]), u \mapsto \boldsymbol{I}_{\tilde\theta}(u)$ in~\eqref{eq:operatorI}, which is performed very efficiently by facilitating transfer learning from other models encountered during the offline training phase. In contrast, when training is done based on noisy in-context examples $(u^{i,1},P^{i,1}), \ldots, (u^{i,M}, P^{i,M})$, the offline learning and online few-shot inference step is no longer a classical operator learning problem. Instead, the object of inference is a map $\boldsymbol{P}_{\tilde\theta}: L^2([0,T]) \rightarrow \mathcal{M}(L^2([0,T])), u \mapsto \boldsymbol{P}_{\tilde\theta} (u)$, where $\mathcal{M}(L^2([0,T]))$ is the set of probability measures on $L^2([0,T])$ and $\boldsymbol{P}_{\tilde\theta} (u)$ denotes the distribution of the trajectory of the execution price $P$ in~\eqref{eq:ExecutionPrice} under model $\tilde\theta$ when trading follows strategy $u$. For a given question condition $u^0$, this map is then queried for the expectation $Y^0_{\cdot} = \mathbb{E}_{\boldsymbol{P}_{\tilde\theta} (u^0)}[P_0-P_{\cdot}]$ in~\eqref{eq:ExecutionPrice} at points $(s_1,\ldots,s_k)$ (assuming that the unaffected price process $S$ in~\eqref{eq:ExecutionPrice} is a martingale), where the expectation is to be understood in a functional sense, i.e., as an element in $L^2([0,T])$. If $S$ is not a martingale, then $Y^0_{\cdot} \neq \mathbb{E}_{\boldsymbol{P}_{\tilde\theta} (u^0)}[P_0-P_{\cdot}]$, and a shifted expectation is learned instead. In general, it is expected that when ICON is trained based on noisy in-context pairs $(u^{i,1}, P^{i,1}), \ldots, (u^{i,M}, P^{i,M})$, the training set and the number of examples $M$ must be considerably larger than when ICON is trained directly based on examples $(u^{i,1},Y^{i,1}), \ldots, (u^{i,M}, Y^{i,M})$ as in Section~\ref{subsec:training:uY:based}. 

\subsection{Neural network solver with ICON surrogate} \label{subsec:OCnet}

We now explain the general methodology for step 2, that is, the OCnet training for the optimal liquidation strategy using ICON as a proxy for the price impact operator. This step is illustrated in Figure~\ref{fig:illustration_icon-oc} (b). More precisely: (i) as described above in Section~\ref{subsec:training:uY:based}, based on $M$ discretized example trajectories $(u^i,Y^i)_{i=1,\ldots,M}$ provided as context, ICON infers a model (here represented by $\theta$) and we obtain an ICON surrogate operator $\boldsymbol{\hat{I}}_{\theta}$, which approximates the true price impact operator $\boldsymbol{I}_{\theta}$ in~\eqref{eq:operatorI}; (ii) we feed $\boldsymbol{\hat{I}}_{\theta}$ into the optimization problem in~\eqref{def:optproblem} and train a neural network policy (OCnet) to approximate the optimal execution strategy in the presence of the current price impact environment detected by ICON. 

Put differently, for a fixed initial inventory $x \in \mathbb{R}$, using the representation in~\eqref{eq:objective2}, the optimization problem becomes 
\begin{equation} \label{eq:optimal_control_surrogate}
\max_{u \in \mathcal{U}}  \, \mathbb{E}\Bigg[ \int_0^T \Big( \big(\alpha_t - (\boldsymbol{\hat{I}}_{\theta}(u))_t \big) u_t  - \varepsilon u_t^2 - \phi X_t^2 \Big) dt - \varrho X_T^2 \Bigg],
\end{equation}
where the in-context transformer network predicts the price impact
\begin{equation} \label{def:ICONsurrogate}
Y_t \approx \boldsymbol{\hat{I}}_{\theta} \left( t, (u_s)_{0 \leq s \leq t};(u^1_s,Y^1_s)_{0 \leq s \leq T},\ldots,(u^M_s,Y^M_s)_{0 \leq s \leq T} \right)\qquad (0 \leq t \leq T)
\end{equation}
incurred by a strategy $u \in \mathcal{U}$. With a slight abuse of notation in~\eqref{def:ICONsurrogate}, we make explicit the path dependence of~$Y_t$ on all past trades $(u_s)_{0 \leq s \leq t}$, as well as the dependence on the $M$ fixed example pairs observed on $[0,T]$ provided as context. 

Next, assuming that trading takes place on an equidistant time grid $t_i = i \Delta t$, $i=0,\ldots,N-1$, with $\Delta t = T/N$, and having access to $L$ samples $(\alpha^{(l)}_{t_i})_{i=0,\ldots,N-1}$, $1 \leq l \leq L$, of the alpha signal $(\alpha_t)_{0 \leq t \leq T}$ in~\eqref{eq:alpha} (modeled by the trader), one can approximate~\eqref{eq:optimal_control_surrogate} via
\begin{align}
& \frac{1}{L} \sum_{l=1}^L \Bigg\{ \sum_{i=0}^{N-1} \bigg( \Big( \alpha_{t_i}^{(l)} - \boldsymbol{\hat{I}}_{\theta} \big(t_i, u_{t_0}, \dots, u_{t_{i-1}}; (u^1_{t_j}, Y^1_{t_j})_{j=0,\ldots,N-1}, \ldots, (u^M_{t_j}, Y^M_{t_j})_{j=0,\ldots,N-1} \big) \Big) u_{t_i} \nonumber \\ 
& \hspace{65pt} - \varepsilon u_{t_i}^2 - \phi X_{t_i}^2 \bigg) \Delta t - \varrho X_{t_{N}}^2 \Bigg\} \rightarrow \max_{u \in \mathcal{N}} \label{eq:optimal_control_surrogate_discrete}
\end{align}
with $X_{t_i} = x - \sum_{j=0}^{i-1} u_{t_j} \Delta t$. Here, $\mathcal{N}$ denotes the set of all feed-forward neural network policies of the form $\text{NN}_{\vartheta}(t,a)$ with $\text{NN}_{\vartheta}: [0,T] \times \mathbb{R} \rightarrow \mathbb{R}$ and some fixed architecture parameterized by $\vartheta \in \mathbb{R}^m$ for some $m \in \mathbb{N}$. Specifically, Remark~\ref{rem:determinstic} motivates in~\eqref{eq:optimal_control_surrogate_discrete} to search for neural network policies $\text{NN}_{\vartheta} \in \mathcal{N}$ as functions in $(t_i,\alpha_{t_i}^{(l)})$, i.e., $u_{t_i} = \text{NN}_{\vartheta}(t_i,\alpha_{t_i}^{(l)})$ and realized alpha signal $\alpha^{(l)}$. The optimal strategy can then be readily computed by employing a policy gradient method; that is, by adopting the approach in~\cite{HanE:16}, running a stochastic gradient descent-type algorithm on the discrete objective functional in~\eqref{eq:optimal_control_surrogate_discrete} with respect to the neural network parameters~$\vartheta$. We refer to the obtained optimal execution strategy as the OCnet policy. More details on the OCnet training are provided in Section~\ref{subsec:OCNETnumerics}. 

\section{Numerical illustrations} \label{sec:numerical}

In this section, we describe our numerical experiments and the results obtained. Specifically, we study our general approach within the broad class of linear propagator models. This class of models was originally developed by~\cite{BouchaudEtAl:04, BouchaudEtAl:09} in discrete time and formulated by~\cite{Gatheral:10} in continuous time; see also~\cite[Chapter 13]{BouchaudEtAl:18}. The associated optimal liquidation problem was explicitly solved only very recently in~\cite{JaberNeuman:25}. We emphasize that the methodology proposed in this paper does not depend on this particular model setup but can be applied to any price impact model specification. We merely focus our numerical study on the linear propagator models because their tractability readily allows us to effectively benchmark our method against explicit optimal solutions (ground truths). 

\begin{example} \label{ex:linearpropagators}
In the class of linear propagator models, the price impact process $Y=(Y_t)_{0 \leq t \leq T}$ in~\eqref{eq:operatorI} is modeled by an integral operator of the form
\begin{equation} \label{eq:Y}
    Y_t = (\boldsymbol{I}_{\theta}(u))_t = \int_0^T G(t-s) 1_{\{s\leq t\}} u_s \lambda \, ds \qquad (0 \leq t \leq T)
\end{equation}
with some price impact kernel $G: [0,T] \rightarrow \mathbb{R}_+$ in $L^2([0,T],\mathbb{R})$ (also called propagator) and constant push factor $\lambda >0$ (also referred to as Kyle’s lambda). In particular, the process $Y$ represents transient price impact that persists and decays over time. The types of parametric kernels $G$ we are considering in~\eqref{eq:Y} are as follows:
\begin{enumerate}
\item[(I)] exponential kernels as proposed by~\cite{ObizhaevaWang:13} of the form
\begin{equation} \label{eq:exponentialkernel}
G(t) = e^{-\beta t} \qquad (0 \leq t \leq T)
\end{equation}
with some impact decay parameter $\beta >0$. In this case, note that $Y$ in~\eqref{eq:Y} is a Markovian process and satisfies the (random) linear ordinary differential equation (ODE)
\begin{equation} \label{eq:exponentialODE}
    Y_0 = 0, \qquad \dot{Y}_t = -\beta Y_t + \lambda u_t \qquad (0 \leq t \leq T).
\end{equation}
\item[(II)] power law kernels as introduced by~\cite{BouchaudEtAl:04, Gatheral:10} of the from
\begin{equation} \label{eq:powerlawkernel}
G(t) = \frac{1}{(\ell + t)^\gamma} \qquad (0 \leq t \leq T)
\end{equation}
for some shift parameter $\ell \geq 0$ and decay elasticity $\gamma>0$. We distinguish between (i) the non-singular case with $\ell > 0$ and $\gamma > 0$, and (ii) the singular case with $\ell = 0$ and $\gamma \in (0,0.5)$. Observe that power law kernels induce non-Markovian dynamics for the impact process $Y$ in~\eqref{eq:Y}. 
\end{enumerate}
Here, we use the subscript $\theta$ in $\boldsymbol{I}_{\theta}$ in~\eqref{eq:Y} to emphasize the dependence of the operator on a certain price impact propagator model with a parametric kernel $G$ of either type (I) or (II) and a push factor~$\lambda$. 
\end{example}

Using the class of propagator models from Example~\ref{ex:linearpropagators} as a testbed, we illustrate the following:
\begin{enumerate}
\item The trained ICON model is capable of accurately inferring the underlying price impact model from only $M=5$ examples provided as context, even for propagator models not present in the training data.
\item Using ICON as a surrogate operator in the optimal execution problem (ICON-OCnet) correctly retrieves the ground truth optimal execution strategies $u^\star$ from~\cite[Proposition 4.5]{JaberNeuman:25} for various propagator models generating the five examples as context.
\end{enumerate}

Throughout our numerical study, we set $T=1$ and perform all training steps for ICON and OCnet, including the in-context examples and queries for the predictions, on a fixed equidistant time grid with $N=100$ steps. Note that this is merely for simplicity and not a requirement. In addition, the hyperparameters in the optimal control problem in~\eqref{eq:optimal_control_surrogate_discrete} are set as follows: we fix the instantaneous cost parameter $\varepsilon$ at $0.5$ and set $\phi = 0$ for the penalty on the running inventory in accordance with the implementation of the ground truth optimal strategies $u^\star$ in~\cite[Section 5]{JaberNeuman:25}, which we use as our benchmark. We also let the penalization on the terminal inventory $\varrho$ be equal to 10 to enforce the liquidation constraint $X_T \approx 0$. 

Moreover, to get a clear picture of the performance of our ICON-OCnet methodology, we focus as in~\cite[Section 3]{NeumanStockingerZhang:23} on the deterministic base case as a testbed and ``neutralize'' the effects of an exogenous intraday alpha signal by assuming that the unaffected price process $S$ in~\eqref{eq:ExecutionPrice} is a martingale. Note that this implies $\alpha \equiv 0$ in~\eqref{eq:alpha}. Hence, without loss of generality, it suffices to optimize in~\eqref{eq:optimal_control_surrogate_discrete} over neural network policies that are functions in time only; recall also Remark~\ref{rem:determinstic}. In particular, we know from the results in~\cite{JaberNeuman:25} that in this case the optimal selling rates $u^\star$ to liquidate an initial positive inventory $x > 0$ are strictly positive, smooth, U-shaped functions in time with varying curvature depending on the propagator kernel $G$ and the push factor~$\lambda$. We refer to~\cite[Section 5.2]{JaberNeuman:25} for more details. 

\subsection{Model parameters} \label{sec:model_parameters}

For our numerical study, we adopt the parameter configuration used in~\cite[Section 3]{NutzWebsterZhao:23}. The time unit is trading days and is denoted by $[T]$. We have $T=1$ so that the strategy trades during one trading day over the interval $[0,1]$ (i.e., 6.5 hours during Nasdaq's opening hours). All trading quantities are expressed as a percentage of the \emph{Average Daily Volume} (ADV\%) and we denote by $[V]$ the volume unit. For instance, a sell order with initial size $x=0.1\,[V]$ represents a sell order of size 10\% ADV. Accordingly, the trading rate $u$ in~\eqref{def:X} is measured in $[V][T]^{-1}$. We focus on a range of $[0.01, 0.20]$ for the initial inventory $x$.

The natural (absolute) unit of $Y$ (and the unaffected price process $S$) in~\eqref{eq:ExecutionPrice} is USD $[\$]$. Ultimately, however, it is more common to think of the price impact $Y$ (and its associated costs) in terms of basis points (bps) relative to some initial benchmark price (e.g., the prevailing decision price $P_0 = S_0$ at the beginning of the trading period). The performance functional in~\eqref{eq:objective2}, respectively~\eqref{eq:optimal_control_surrogate_discrete}, is then also standardized accordingly to represent total implementation shortfall costs in basis points. Therefore, in the units of the parameters below, we treat $Y$ as unitless (i.e., standardized). For the kernel parameters in (I) and (II), and the push factor $\lambda$ in Example~\ref{ex:linearpropagators}, we focus on the following ranges: 
\begin{itemize}
\item[(i)] The impact decay $\beta$ in~\eqref{eq:exponentialODE} has unit $[T]^{-1}$ and describes the speed at which impact reverts to zero. The corresponding half-life is given by $\log(2)/\beta$. We let $\beta$ vary in the interval $[0.462, 9.011]$, which roughly corresponds to half-lives ranging from 30 minutes to 1.5 days.
\item[(ii)] The push factor $\lambda$ in~\eqref{eq:Y} has unit $[V]^{-1}$ with values in $[0.1,0.5]$.
\item[(iii)] The power law kernel in~\eqref{eq:powerlawkernel} captures multiple timescales of decay in a parsimonious way. In the non-singular case, we set $\ell = 1$ and let $\gamma$ vary in $[0.3, 1.5]$. For the singular case $\ell =0$, we restrict $\gamma$ to $[0.35,0.45]$.
\end{itemize}


\subsection{Data generation} \label{subsec:data}

\begin{figure}[htbp]
    \centering
    \begin{subfigure}{0.4\textwidth}
        \centering \includegraphics[width=\textwidth]{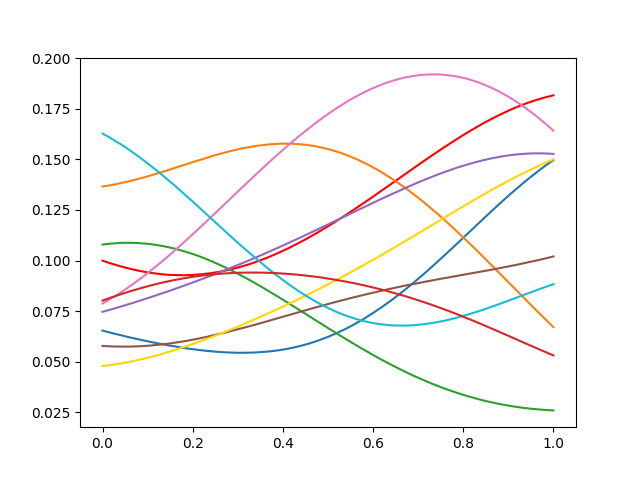}
        \caption{selling rates $u$}
    \end{subfigure}
    \begin{subfigure}{0.4\textwidth}
        \centering \includegraphics[width=\textwidth]{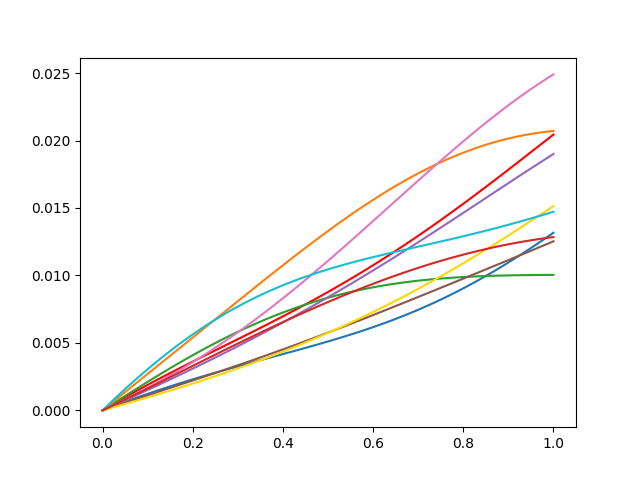}
        \caption{exponential $Y$}
    \end{subfigure}  
    \begin{subfigure}{0.4\textwidth}
        \centering \includegraphics[width=\textwidth]{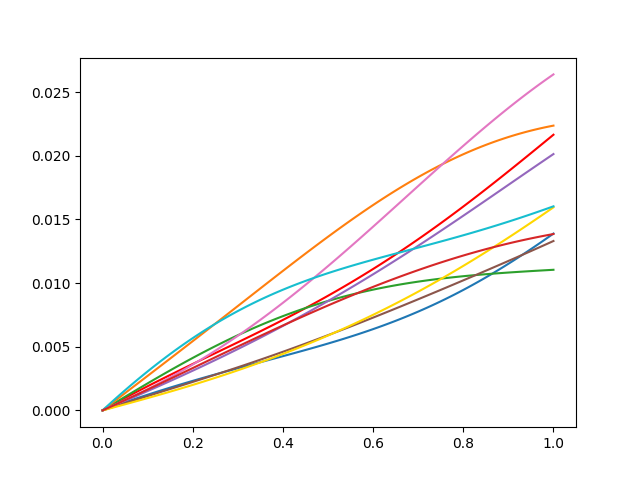}
        \caption{non-singular $Y$}
    \end{subfigure}
    \begin{subfigure}{0.4\textwidth}
        \centering \includegraphics[width=\textwidth]{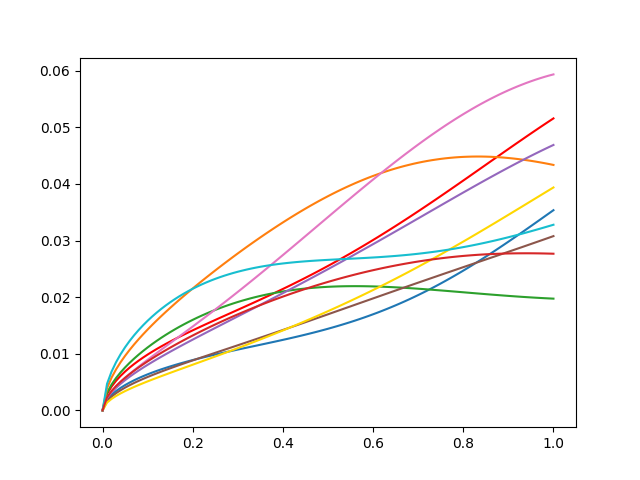}
        \caption{singular $Y$}
    \end{subfigure}
    \caption{Illustration of 10 training trajectories (selling rates $u$ and corresponding price impact $Y$) for the three different kernels with parameters $\lambda = 0.2$, $\beta = 0.5$, $\gamma = 0.45$.}
    \label{fig:data}
\end{figure}

We follow the training procedure outlined in Section \ref{subsec:training:uY:based} based on in-context pairs $(u,Y)$. For the offline training step of ICON (recall Figure~\ref{fig:illustration_icon-oc}~(a)), we generate three different synthetic datasets of selling rates and price impact trajectories $(u,Y)$ corresponding to the three different kernels in~\eqref{eq:exponentialkernel} and~\eqref{eq:powerlawkernel}: exponential, non-singular power law, singular power law. Here, we take the exponential kernel in~\eqref{eq:exponentialkernel} as an example to explain our training methodology. First, we randomly sample 80,000 pairs of hyperparameters $\theta = (\lambda,\beta)$ for the push factor $\lambda \sim U([0.1, 0.5])$ and the impact decay $\beta \sim U([0.462, 9.011])$; cf.~also Section~\ref{sec:model_parameters}. Next, for each $\theta$, we generate 10 strictly positive selling rates $u = \tilde{u} + 0.1$, where $\tilde{u}$ is sampled from a smooth Gaussian process on $[0, 1]$ with kernel $0.05^2 \exp(-2t^2)$. Recall from Section~\ref{sec:model_parameters} that the selling rate is measured in $[V][T]^{-1}$ with ADV as volume unit. We then compute for each $u$ the corresponding price impact $Y = \boldsymbol{I}_{\theta}(u)$ using a discretized version of~\eqref{eq:Y}. The other two datasets for the power law kernels in~\eqref{eq:powerlawkernel} are generated similarly, with hyperparameters $\lambda \sim U([0.1, 0.5])$, $\gamma \sim U([0.3, 1.5])$ and $\ell = 1$ for the non-singular kernel; and $\lambda \sim U([0.1, 0.5])$, $\gamma \sim U([0.35, 0.45])$ for the singular kernel with $\ell = 0$. An illustration of 10 sample trajectories used in training is shown in Figure~\ref{fig:data}.

During each iteration in the training process of ICON, a mini-batch of size 8 of hyperparameters $\theta_i$, $i=1,\ldots,8$, is randomly selected from the 80,000 generated data points. Then, for each such $\theta_i$ with associated 10 generated sample pairs $((u^{i,1}, Y^{i,1}),\ldots,(u^{i,10}, Y^{i,10}))$ as described above, we randomly select one pair as the input-output pair (i.e., question condition $u^{i,0}$ and labeled output $Y^{i,0}$ as illustrated in Figure~\ref{fig:illustration_icon-oc}~(a)); and from the remaining 9 pairs we randomly select $M=5$ to serve as examples in the context. A stochastic gradient descent step (using scheduled AdamW) on the training loss function (i.e., update on the ICON's parameters) is then performed with this mini-batch of 8 data points. The loss function is a standard mean squared error of the prediction on the time grid on $[0,1]$, referred to as the $\ell^2$ error.


\subsection{ICON performance}\label{subsec:ICONnumerics}

\begin{figure}[htbp]
    \centering 
    \includegraphics[width=.6\textwidth]{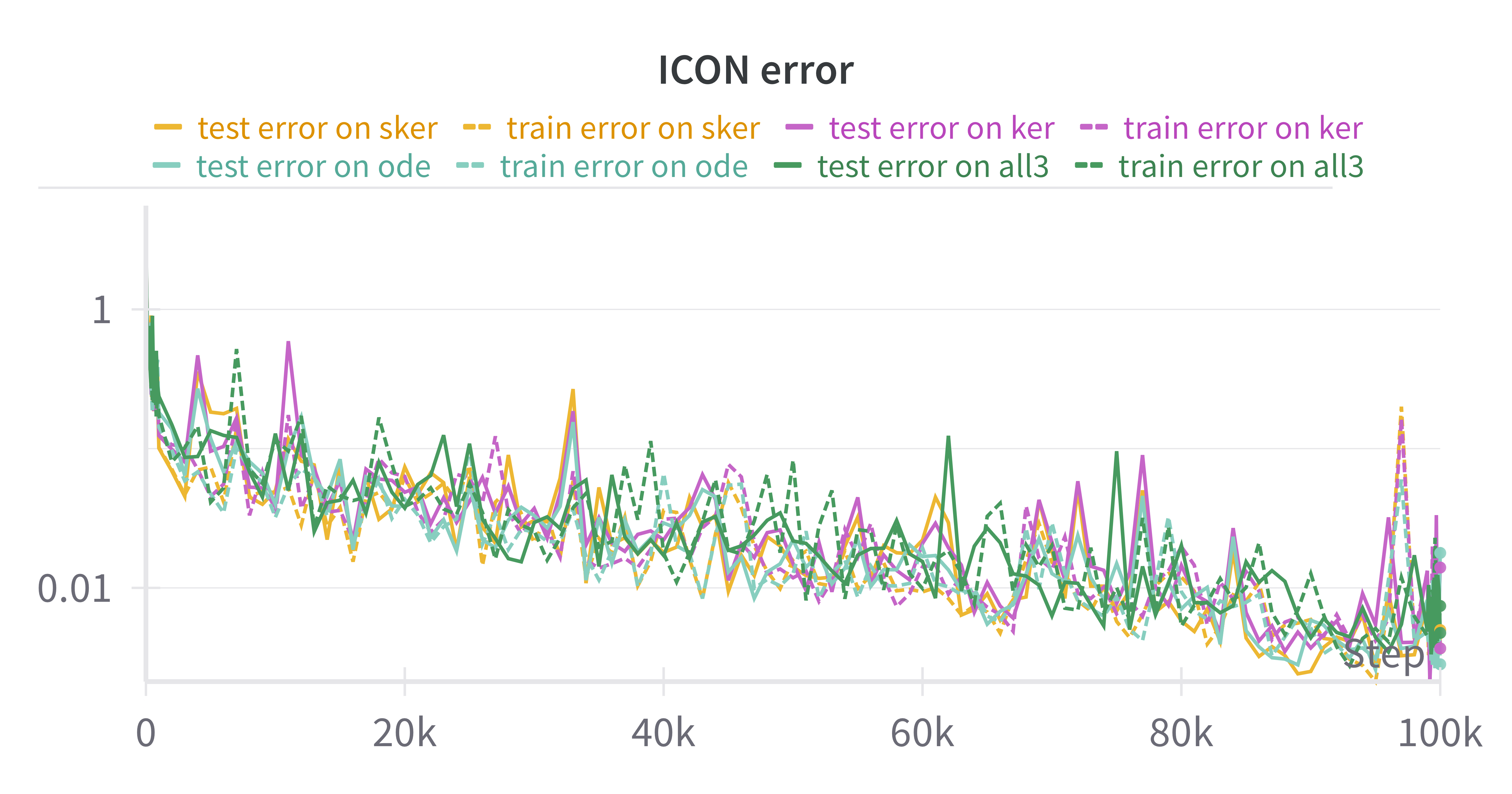}
    \caption{ICON error (on the training set and a separate test set) versus training iterations with 100,000 training steps in total.}
    \label{fig:ICON_training_err}
\end{figure}

In this section, we evaluate the performance of ICON training described in Section~\ref{subsec:training:uY:based}. In total, four different ICON models are trained: (i) three models, each trained on a separate propagator model dataset \texttt{ode}, \texttt{ker}, \texttt{sker} generated from the three different propagator kernels exponential (\texttt{ode}), non-singular power law (\texttt{ker}), singular power law (\texttt{sker}) as described above in Section~\ref{subsec:data}; (ii) a fourth model trained on a mixed dataset \texttt{all3} that combines the previous three datasets. To ensure a fair comparison, the number of training samples is 80,000 in all four datasets (i.e., in \texttt{all3} only 1/3 of each dataset \texttt{ode}, \texttt{ker}, \texttt{sker} of size 80,000 is used). Inspired by~\cite{YANG2025107455} we use an encoder-only transformer architecture with 6 layers, 8 heads, head dimension 256, model dimension 256, and a widening factor of 4. We train all four ICON models for 100,000 iterations, roughly corresponding to 10 epochs. Training is performed on a single NVIDIA GeForce RTX 4090 GPU and takes about 3.5 hours. We refer the reader to~\cite{NIPS2017_3f5ee243} where this architecture was first introduced, and~\cite{Alammar2018IllustratedTransformer} for a non-technical illustration. The resulting~$\ell^2$ relative error is plotted in Figure~\ref{fig:ICON_training_err} against the number of iterations.

After training, we assess the relative $\ell^2$ errors of all four ICON model predictions $\boldsymbol{\hat{I}}_\theta(u)$ with respect to the exact output $Y=\boldsymbol{I}_\theta(u)$ of the provided question condition $u$, for various in-context examples from propagator models $\theta$ from out-of-sample test datasets. In particular, we examine the out-of-distribution performance (i.e., transfer learning) of the first three ICON models (trained separately on \texttt{ode}, \texttt{ker}, \texttt{sker}, respectively) on in-context examples from a propagator kernel type not seen during the training phase. The test datasets, denoted by $\texttt{ode\_t}$, \texttt{ker\_t}, \texttt{sker\_t}, are generated using the exact same methodology as for the training data described in Section~\ref{subsec:data}. The results are summarized in Table~\ref{tab:icon_err}. Each column represents a trained ICON model and each row corresponds to a specific test dataset for the in-context examples. For each combination, we calculate the mean and standard deviation of the relative $\ell^2$ prediction error for a sample of size $576$. The in-distribution errors (bold values in the diagonal of the first three columns) are very small for each ICON model, illustrating that the trained ICON models are capable of accurately inferring the underlying propagator model from the $M=5$ in-context examples. Moreover, also the out-of-distribution errors for the transfer learning are quite small (off-diagonal entries of the first three columns). Hence, the ICON prediction based on few-shot learning from the prompted examples generalizes fairly well to unseen propagator kernel types. We also observe that the ICON model trained on the mixed dataset \texttt{all3} performs very well across all three in-context example training sets.     

\begin{table}[htbp]
    \centering
    \begin{tabular}{c|c|c|c|c}
        \hline
        \hline
        & \texttt{ode} & \texttt{ker} & \texttt{sker} & \texttt{all3}\\
        \hline 
        \texttt{ode\_t} & $\mathbf{0.0053 \pm 0.0045}$ &$0.1075 \pm 0.1266$ &$0.1635 \pm 0.2369$ &$0.0060 \pm 0.0044$ \\
        \hline
        \texttt{ker\_t} & $0.0072 \pm 0.0057$ &$\mathbf{0.0045 \pm 0.0024}$ &$0.0345 \pm 0.0309$ &$0.0063 \pm 0.0048$ \\
        \hline
        \texttt{sker\_t} & $0.0423 \pm 0.0191$ &$0.0392 \pm 0.0241$ &$\mathbf{0.0052 \pm 0.0036}$ & $0.0057 \pm 0.0036$ \\
        \hline
        \hline
    \end{tabular}
    \caption{ICON error table of relative $\ell^2$ prediction errors. Each column corresponds to one of the four ICON models trained on a specific dataset. The rows represent the different test datasets for the 5 in-context examples.} 
    \label{tab:icon_err}
\end{table}

\begin{figure}[htbp]
    \centering
    \begin{subfigure}{0.35\textwidth}
        \centering \includegraphics[width=\textwidth]{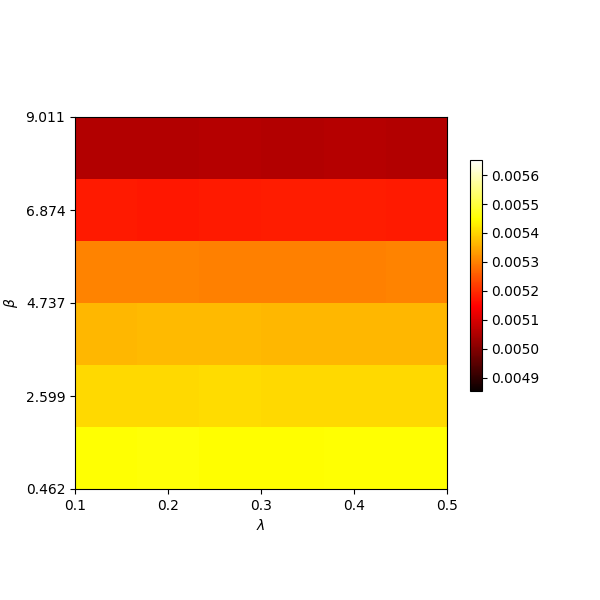}
        \caption{\texttt{ode\_t}, \texttt{ode}}
    \end{subfigure}
    \begin{subfigure}{0.35\textwidth}
        \centering \includegraphics[width=\textwidth]{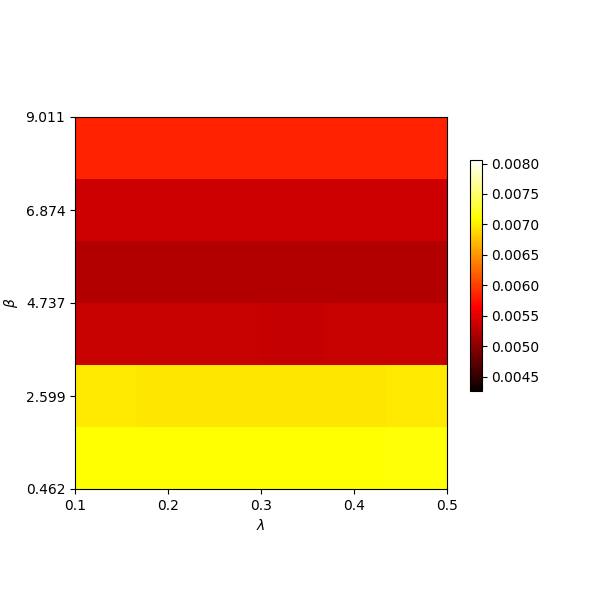}
        \caption{\texttt{ode\_t}, \texttt{all3}}
    \end{subfigure}
    \begin{subfigure}{0.35\textwidth}
        \centering \includegraphics[width=\textwidth]{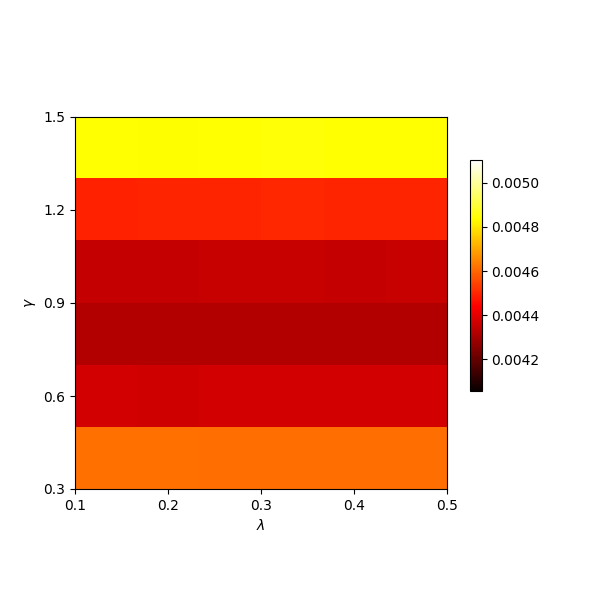}
        \caption{\texttt{ker\_t}, \texttt{ker}}
    \end{subfigure}
    \begin{subfigure}{0.35\textwidth}
        \centering \includegraphics[width=\textwidth]{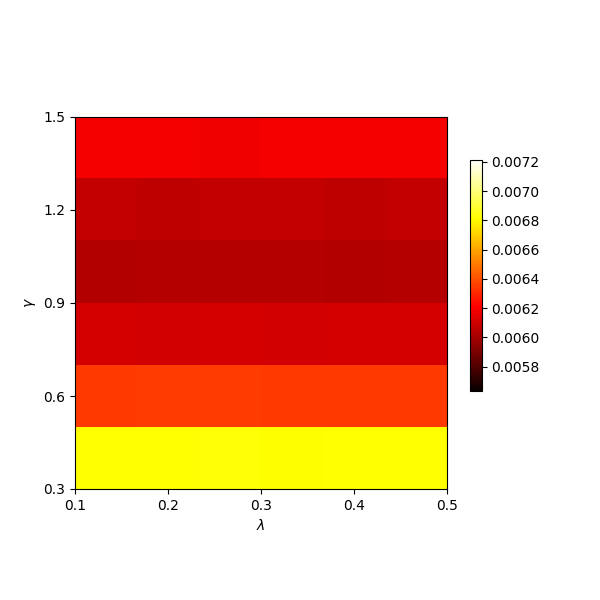}
        \caption{\texttt{ker\_t}, \texttt{all3}}
    \end{subfigure}
    \begin{subfigure}{0.35\textwidth}
        \centering \includegraphics[width=\textwidth]{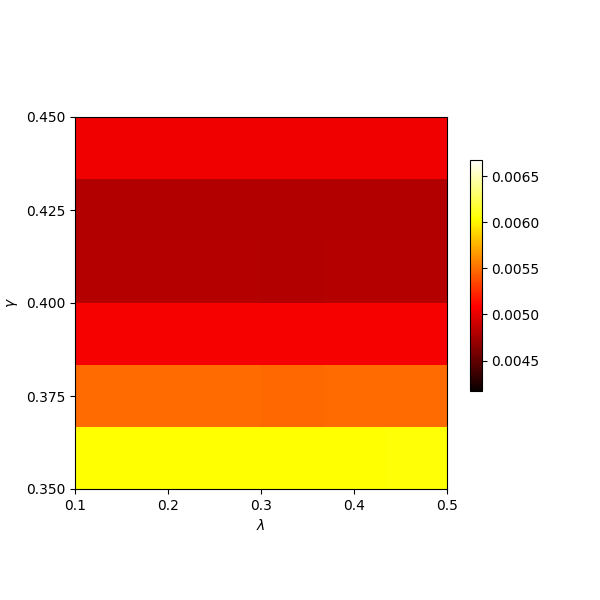}
        \caption{\texttt{sker\_t}, \texttt{sker}}
    \end{subfigure} 
    \begin{subfigure}{0.35\textwidth}
        \centering \includegraphics[width=\textwidth]{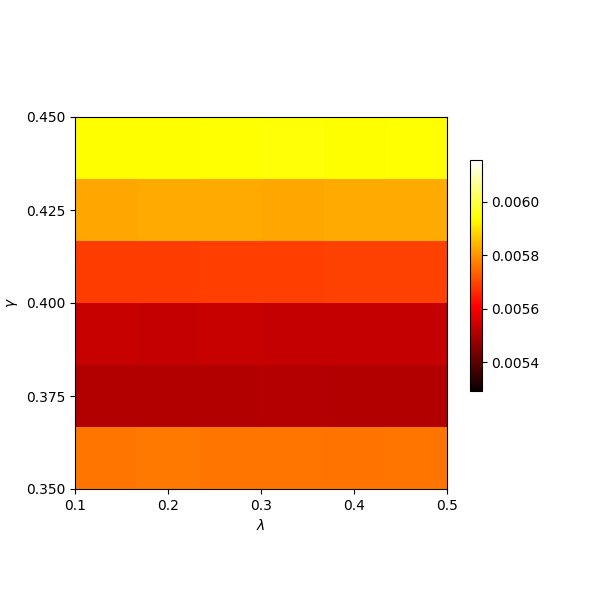}
        \caption{\texttt{sker\_t}, \texttt{all3}}
    \end{subfigure}
    \caption{Heatmaps for ICON in-distribution errors for different types of in-context examples (first label) and ICON models trained on a specific dataset (second label). The value of each box represents the mean error over 16 random samples of sets of hyperparameters $\theta$ from the corresponding ranges, generating the five in-context examples ($x$-axis represents values for $\lambda$, $y$-axis represents values for $\beta$ and $\gamma$, respectively).}
    \label{fig:icon_heatmap_id}
\end{figure}

Next, we present in Figure~\ref{fig:icon_heatmap_id} heatmaps which shed some light on the dependence of the ICON models' in-distribution prediction errors with respect to the underlying price impact model parameters generating the in-context examples. More precisely, the two axes represent the hyperparameters $\theta$ for each kernel type (I) and (II) (split in 6 intervals), with the color of each box indicating the mean error of the ICON prediction for 16 randomly selected hyperparameters within the box's range  which are used for generating the in-context example pairs and the exact output label $\boldsymbol{I}_\theta(u)$ of the question condition $u$ (again, everything is generated using the same method as described in Section~\ref{subsec:data}). Note that the error remains invariant with respect to the push factor $\lambda$ due to the scaling-invariant property of our ICON architecture. With regards to the impact decay $\beta$ in the exponential kernel, the error becomes smaller with faster decay (Figure~\ref{fig:icon_heatmap_id}, (a) and (b)). For the power law kernel, the influence of $\gamma$ shows a slightly less coherent pattern across the different ICON models \texttt{ker}, \texttt{sker}, \texttt{all3} (Figure~\ref{fig:icon_heatmap_id}, (c)--(f)); perhaps not surprisingly due to the more subtle nature of a power law decay. Overall, however, the errors are very small.

\begin{figure}[htbp]
    \centering
    \begin{subfigure}{0.35\textwidth}
        \centering \includegraphics[width=\textwidth]{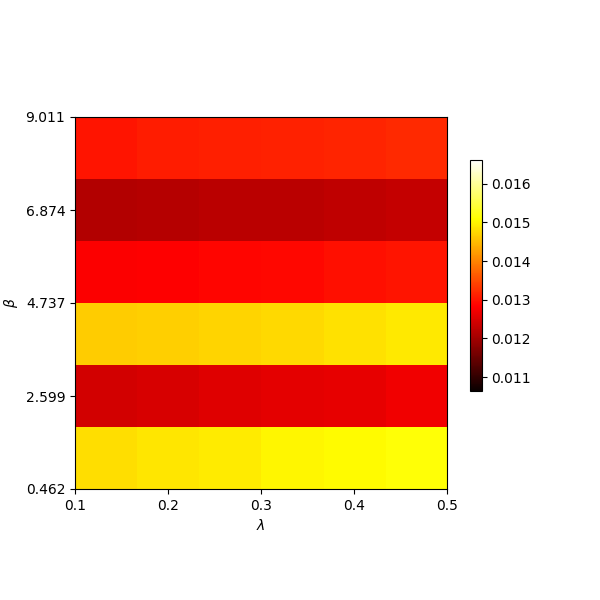}
        \caption{\texttt{ode\_t}, \texttt{ode}}
    \end{subfigure}
    \begin{subfigure}{0.35\textwidth}
        \centering \includegraphics[width=\textwidth]{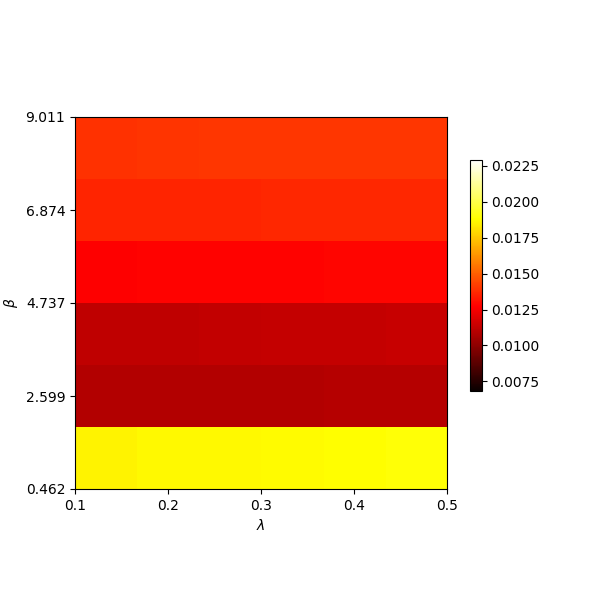}
        \caption{\texttt{ode\_t}, \texttt{all3}}
    \end{subfigure}
    \begin{subfigure}{0.35\textwidth}
        \centering \includegraphics[width=\textwidth]{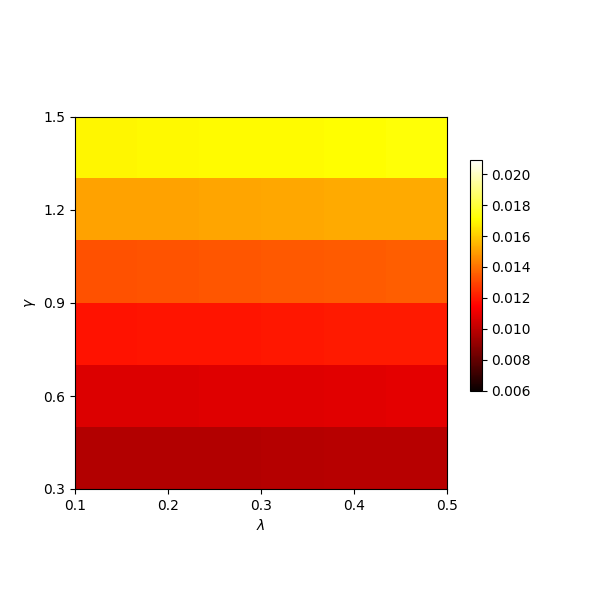}
        \caption{\texttt{ker\_t}, \texttt{ker}}
    \end{subfigure}
    \begin{subfigure}{0.35\textwidth}
        \centering \includegraphics[width=\textwidth]{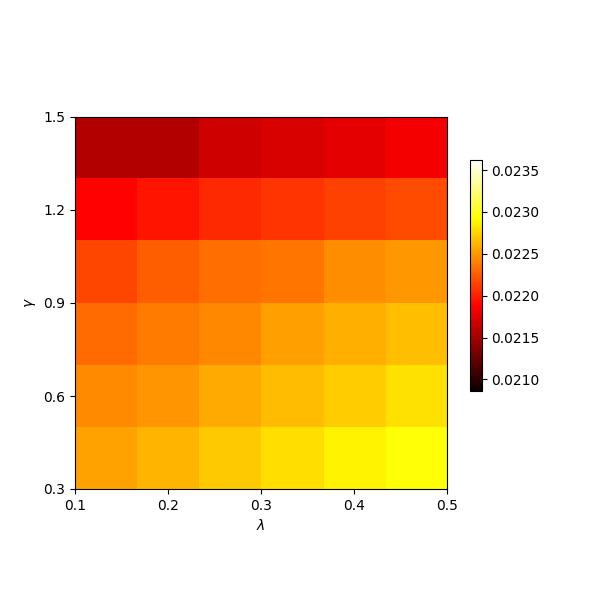}
        \caption{\texttt{ker\_t}, \texttt{\texttt{all3}}}
    \end{subfigure}
    \begin{subfigure}{0.35\textwidth}
        \centering \includegraphics[width=\textwidth]{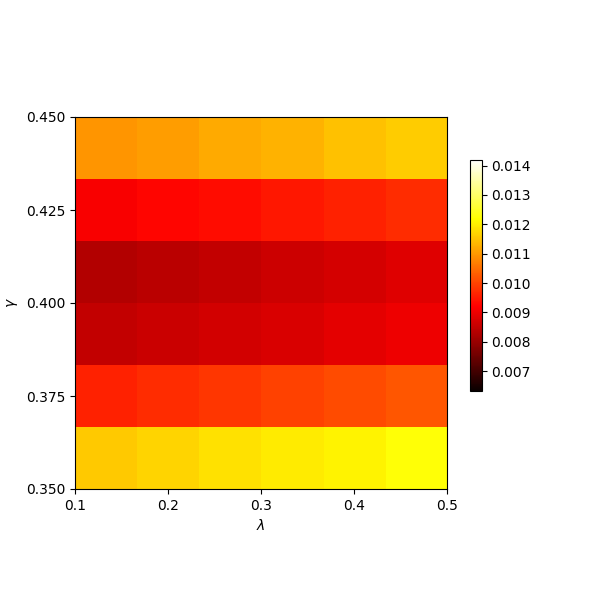}
        \caption{\texttt{sker\_t}, \texttt{sker}}
    \end{subfigure} 
    \begin{subfigure}{0.35\textwidth}
        \centering \includegraphics[width=\textwidth]{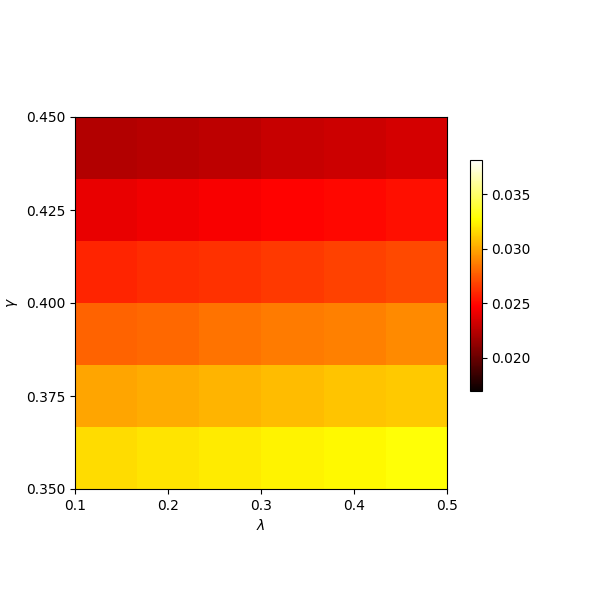}
        \caption{\texttt{sker\_t}, \texttt{all3}}
    \end{subfigure}
    \caption{Heatmaps similar to Figure~\ref{fig:icon_heatmap_id} for ICON errors for different types of in-distribution in-context examples (first label) and ICON models trained on a specific dataset (second label). Here, the question condition for the ICON prediction is the out-of-distribution optimal execution strategy $u^\star$ of the corresponding propagator model associated with hyperparameter $\theta$. The value of each box represents the mean error over 16 random samples of sets of hyperparameters $\theta$ from the corresponding ranges ($x$-axis represents values for $\lambda$, $y$-axis represents values for $\beta$ and $\gamma$, respectively).}
    \label{fig:icon_heatmap_ood}
\end{figure}

\begin{figure}[htbp]
    \centering
    \begin{subfigure}{0.5\textwidth}
        \centering \includegraphics[width=\textwidth]{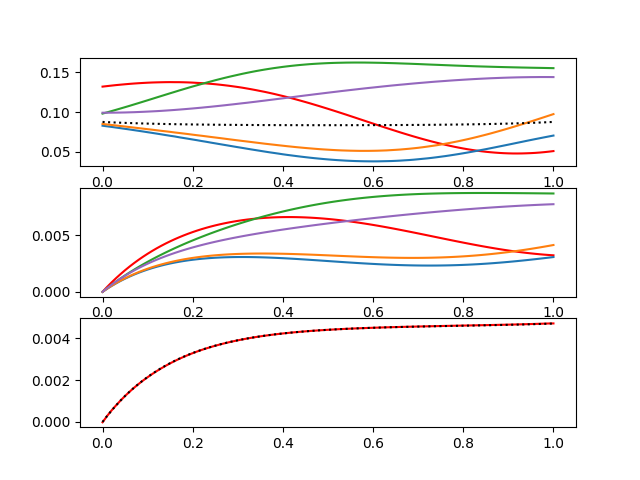}
        \caption{exponential kernel}
    \end{subfigure}
    \begin{subfigure}{0.5\textwidth}
        \centering \includegraphics[width=\textwidth]{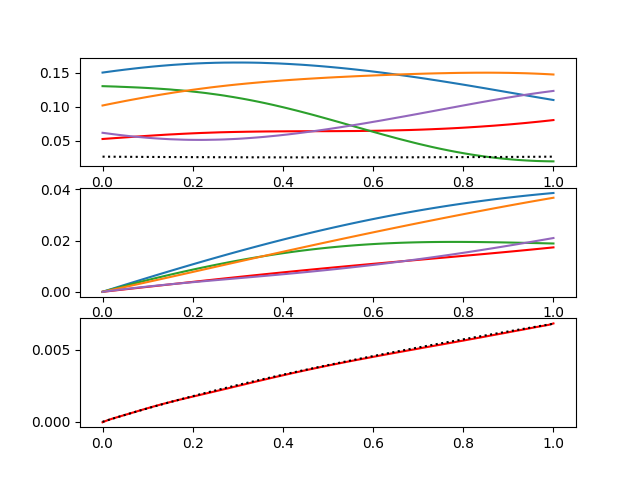}
        \caption{non-singular power law kernel}
    \end{subfigure}
    \begin{subfigure}{0.5\textwidth}
        \centering \includegraphics[width=\textwidth]{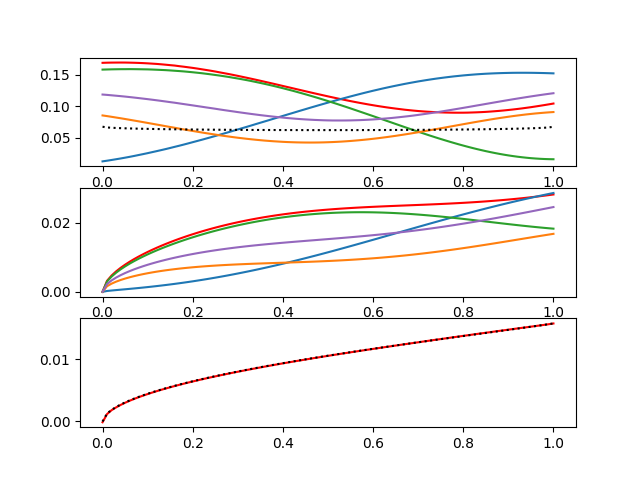}
        \caption{singular power law kernel}
    \end{subfigure}
    \caption{Illustration of the out-of-distribution prediction for $i=1,2,3$ different ICON models trained on \texttt{ode} (a), \texttt{ker} (b), \texttt{sker} (c), respectively. \emph{Top panels:} Five selling rates $u^{i,1},\ldots,u^{i,5}$ as in-context example conditions (colored solid lines) together with the optimal execution strategy $u^{i,\star}$ (black dashed line). \emph{Middle panels:} The five corresponding in-context example price impact trajectories $Y^{i,1}=\boldsymbol{I}_{\theta_i}(u^{i,1}),\ldots,Y^{i,5}=\boldsymbol{I}_{\theta_i}(u^{i,5})$. \emph{Bottom panels:} The ICON surrogate operator prediction $\boldsymbol{\hat{I}}_{\theta_i}(u^{i,\star})$ (black dotted line) compared to the ground truth optimal price impact $Y^{i,\star} = \boldsymbol{I}_{\theta_i}(u^{i,\star})$ (red solid line).}
    \label{fig:icon_gtu_visualization}
\end{figure}

Finally, we evaluate how well ICON generalizes to \emph{unseen} selling rates $u$ as question conditions, that is, strategies which are not directly generated from the Gaussian process as in the synthetic training datasets. This is a first sanity check of ICON's effectiveness as a surrogate operator in the downstream optimal execution problem in~\eqref{eq:optimal_control_surrogate_discrete}, i.e., our ICON-OCnet approach. To this end, we generate five synthetic examples as context from a specific propagator model $\theta$ using the same method as described in Section~\ref{subsec:data}. However, for the in-context prediction task, we use as question condition the actual (out-of-distribution) ground truth optimal execution strategy $u^\star$ of model $\theta$ which is computed via the implementation provided by~\cite{JaberNeuman:25}. We then compute the corresponding optimal price impact process $Y^\star = \boldsymbol{I}_\theta(u^\star)$ to obtain the exact output label, and compare it to the prediction of the ICON surrogate operator $\boldsymbol{\hat{I}}_\theta(u^\star)$ acting on $u^\star$. The relative $\ell^2$ error under this setup is computed and visualized as heatmaps with varying hyperparameters in Figure~\ref{fig:icon_heatmap_ood}, similar to the ones in Figure~\ref{fig:icon_heatmap_id}. This time, the error increases slightly, but still remains fairly small across the different ICON models and test datasets providing the context. We also observe that here, different compared to Figure~\ref{fig:icon_heatmap_id}, the errors depend on the push factor~$\lambda$, simply because variations in $\lambda$ alter the question condition $u^\star$.  

To further elaborate on the previous analysis, we illustrate in Figure~\ref{fig:icon_gtu_visualization} five pairs of in-context examples $((u^{i,1},Y^{i,1}),\ldots,(u^{i,5},Y^{i,5}))$ for $i=1,2,3$ different propagator models with randomly sampled hyperparameters $\theta_i$, generated as described in Section~\ref{subsec:data}; the question condition $u^{i,\star}$ (i.e., the optimal execution strategy for model $\theta_i$); and the prediction curves $\boldsymbol{\hat{I}}_{\theta_i}(u^{i,\star})$ compared to the ground truth price impact $\boldsymbol{I}_{\theta_i}(u^{i,\star})$. More precisely, each subfigure in Figure~\ref{fig:icon_gtu_visualization} corresponds to a specific model $\theta_i$ (exponential kernel in (a), non-singular power law kernel in (b), singular power law kernel in (c)). The top panel of each subfigure displays the in-context example conditions $u^{i,1}, \ldots, u^{i,5}$ in colored solid lines, as well as the question condition $u^{i,\star}$ in black dotted lines. The middle panel shows the corresponding in-context example quantities of interest $Y^{i,1}=\boldsymbol{I}_{\theta_i}(u^{i,1}), \ldots, Y^{i,5}=\boldsymbol{I}_{\theta_i}(u^{i,5})$. The bottom panel compares the in-context prediction $\boldsymbol{\hat{I}}_{\theta_i}(u^{i,\star})$ (red solid line) of the ICON model (trained on the dataset \texttt{ode} in (a), \texttt{ker} in (b), \texttt{sker} in (c), respectively) with the ground truth $\boldsymbol{I}_{\theta_i}(u^{i,\star})$ (black dotted line). Note that the optimal execution strategies $u^{i,\star}$ are almost flat for the three different propagator kernels with (randomly picked) hyperparameters~$\theta_i$. Remarkably, we observe that the three ICON surrogate operators are indeed capable of accurately predicting the corresponding price impact incurred by the selling rates $u^{1,\star}, u^{2,\star}, u^{3,\star}$, which gradually build up at different speeds and magnitudes, depending on the different propagator kernels and their hyperparameters $\theta_1 = (\beta_1, \lambda_1)$ (for the exponential kernel), $\theta_2 = (\gamma_1, \lambda_2)$ (for the non-singular power law kernel) and $\theta_3 = (\gamma_2, \lambda_3)$ (for the singular power law kernel), all of which are correctly inferred by the ICON models from the five prompted in-context examples.

\subsection{ICON-OCnet performance} \label{subsec:OCNETnumerics}

We now illustrate the performance of our ICON-OCnet method described in Section~\ref{subsec:OCnet}. Specifically, using a pretrained ICON model as described above (trained on \texttt{ode}, \texttt{ker}, \texttt{sker}, \texttt{all3}, respectively, and prompted with 5 in-context examples from a specific propagator model $\theta$) as a surrogate operator $\boldsymbol{\hat{I}}_\theta$ in the optimal execution problem, we perform a policy gradient method  directly on the objective function in~\eqref{eq:optimal_control_surrogate_discrete}, and train a simple feed-forward neural network\footnote{The OCnet neural network structure consists of 2 hidden layers, 128 nodes in each layer, and GELU activation function.} with 60,000 iterations for the optimal order execution task. The initial inventory $x$ expressed in ADV is sampled from a uniform distribution $U([0.01,0.2])$. We compare the obtained OCnet policy $\hat{u}$ (with corresponding inventory~$\hat{X}$ and price impact process~$\hat{Y}$) to the actual ground truth optimal execution strategy $u^\star$ (with state processes $X^\star, Y^\star$) of the propagator model $\theta$ generating the in-context examples, which is computed via the implementation provided by~\cite{JaberNeuman:25}. 

\begin{figure}[htbp]
    \centering
    \begin{subfigure}{0.3\textwidth}
        \centering \includegraphics[width=\textwidth]{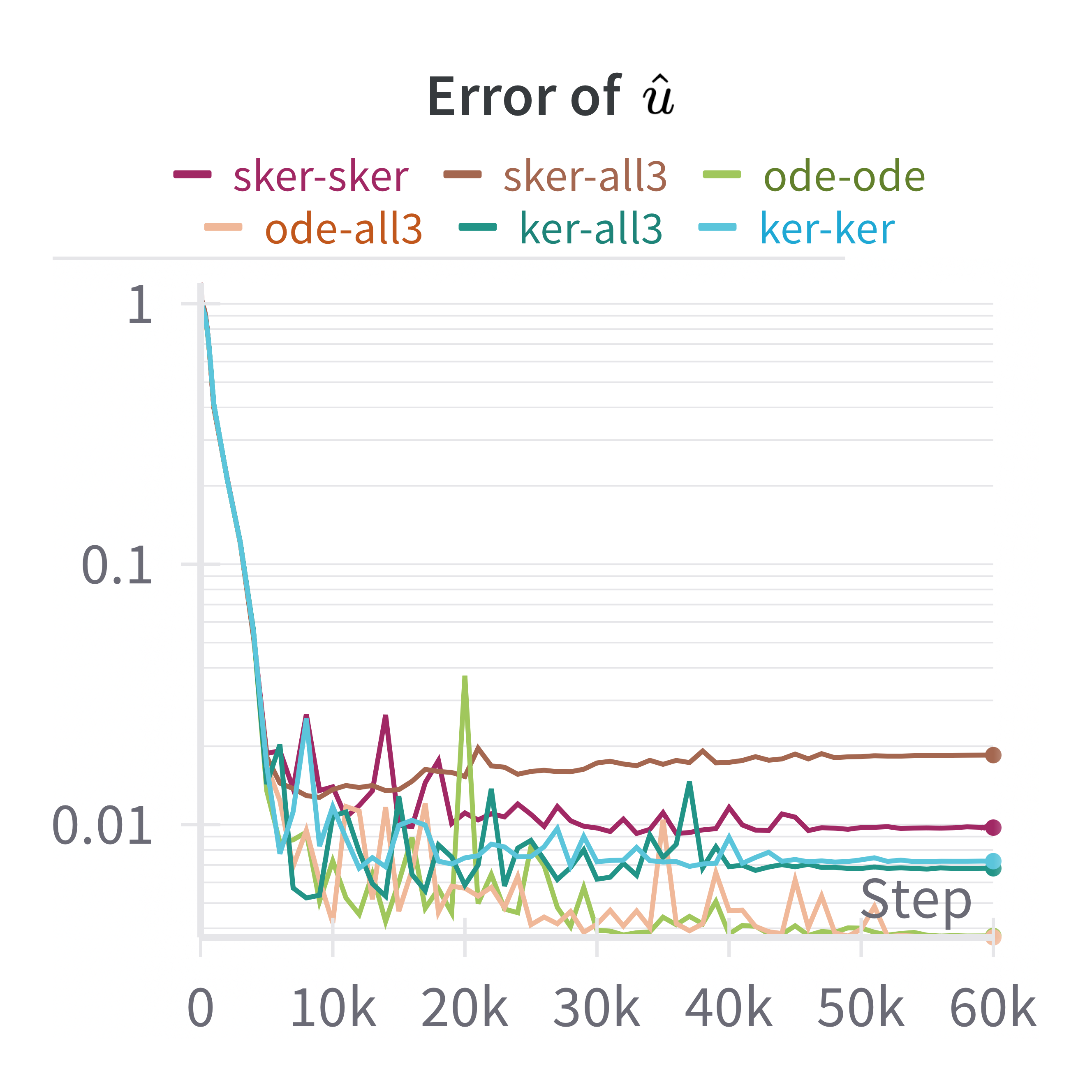}
        \caption{selling rate $\hat{u}$}
    \end{subfigure}
    \begin{subfigure}{0.3\textwidth}
        \centering \includegraphics[width=\textwidth]{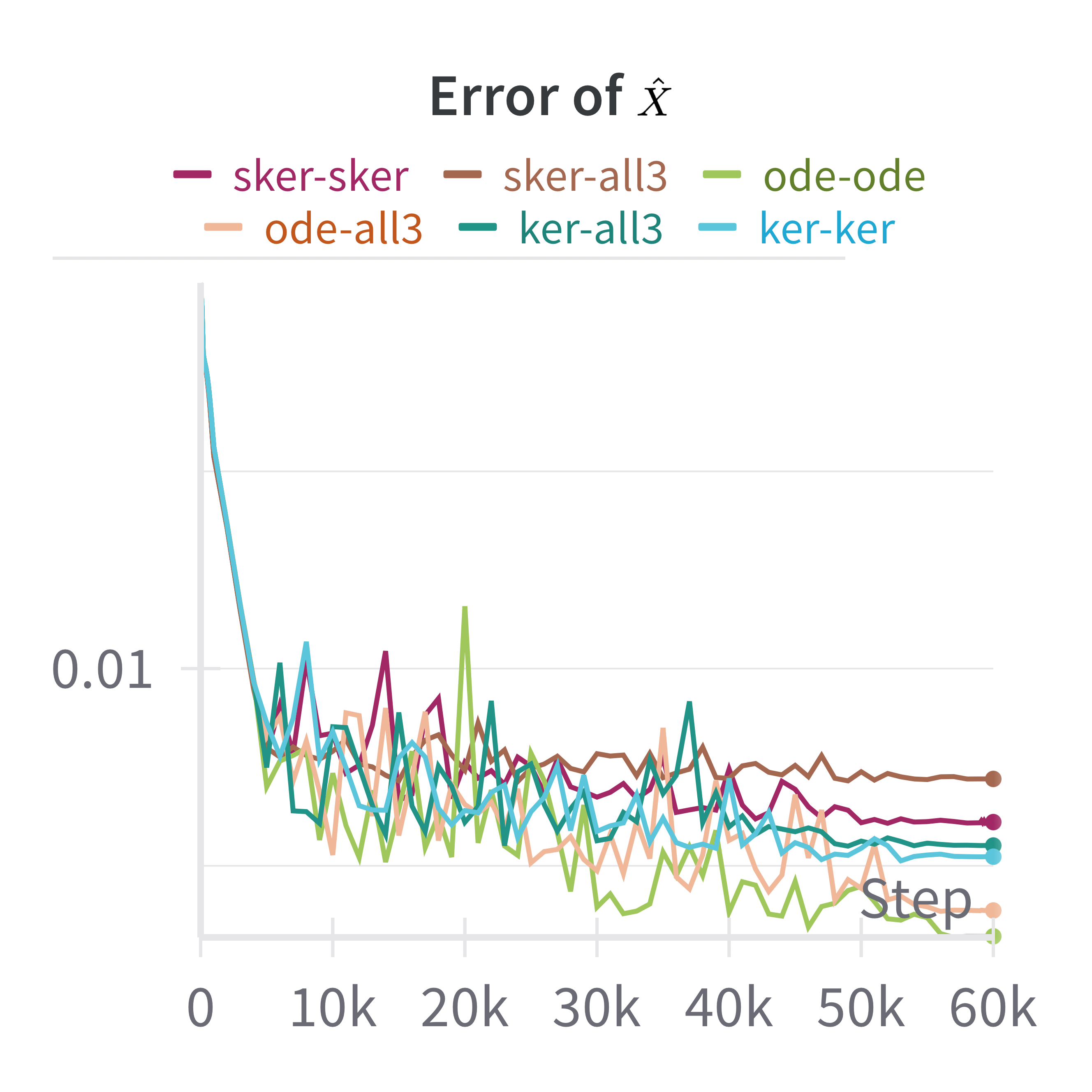}
        \caption{inventory $\hat{X}$}
    \end{subfigure}
    \begin{subfigure}{0.3\textwidth}
        \centering \includegraphics[width=\textwidth]{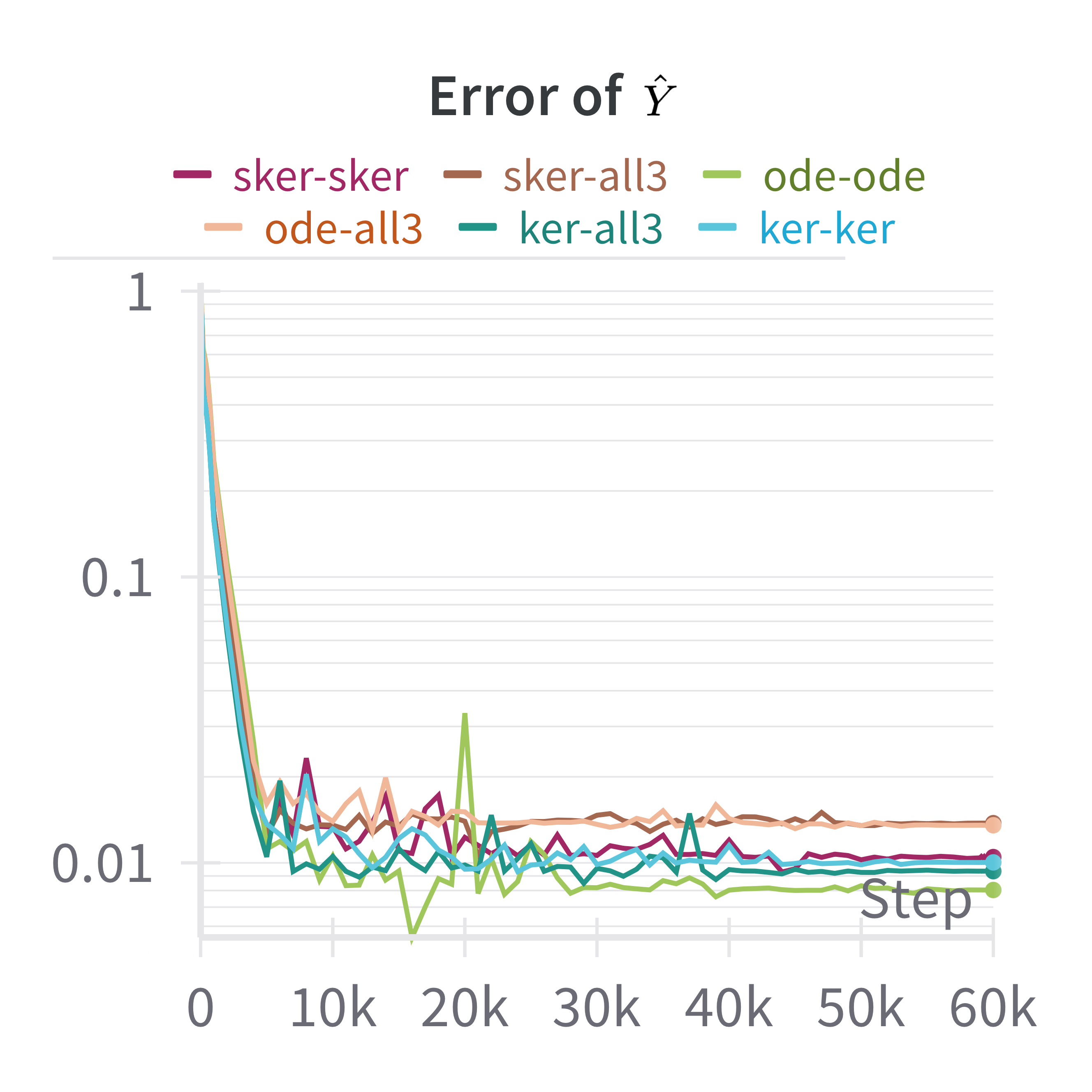}
        \caption{price impact $\hat{Y}$}
    \end{subfigure}
    \caption{Relative errors of $\hat{u}, \hat{X},\hat{Y}$ versus training iterations for different ICON models (second label) used as surrogate operator in the optimal execution problem with a specific propagator model $\theta$ providing the in-context examples (first label). The error values are mean values over 16 randomly selected hyperparameters $\theta$ and initial inventories $x$.
    }
    \label{fig:err_oc_training}
\end{figure}

The relative $\ell^2$ errors of selling rate, inventory, and price impact process (compared with the ground truth) versus the number of training steps are illustrated in Figure~\ref{fig:err_oc_training}. We observe convergence of the policy gradient method after 60,000 iterations. In fact, the errors already drop to an acceptable level after 20,000 steps.    

\begin{figure}[htbp]
    \centering
    \begin{subfigure}{0.6\textwidth}
        \centering \includegraphics[width=\textwidth]{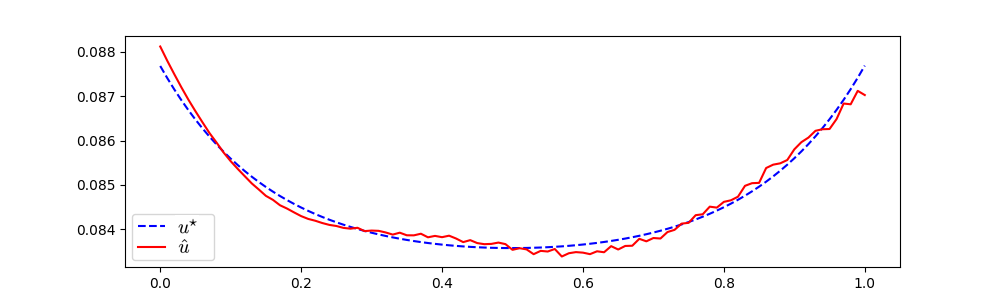}
        \centering \includegraphics[width=\textwidth]{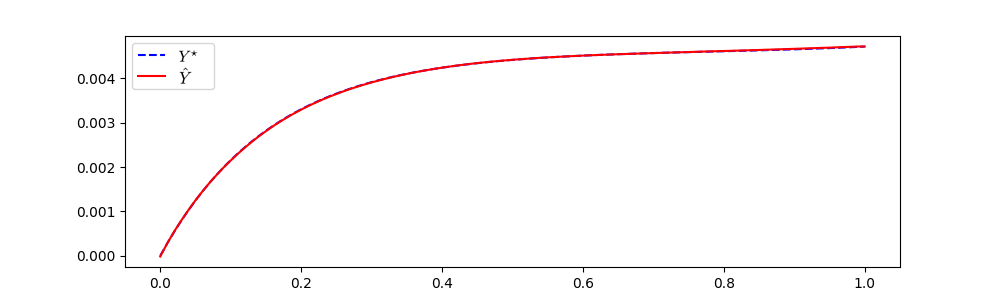}
        \caption{exponential kernel}
    \end{subfigure}
    \begin{subfigure}{0.6\textwidth}
        \centering \includegraphics[width=\textwidth]{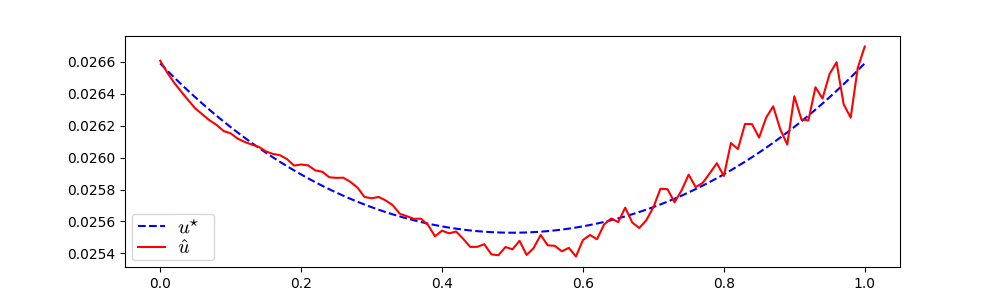}
        \centering \includegraphics[width=\textwidth]{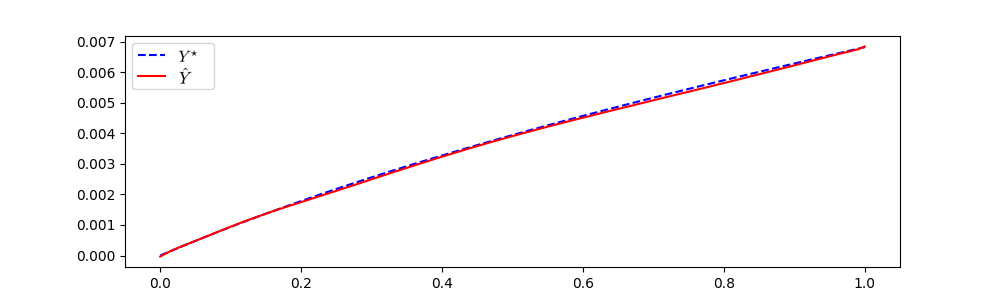}
        \caption{non-singular power law kernel}
    \end{subfigure}
    \begin{subfigure}{0.6\textwidth}
        \centering \includegraphics[width=\textwidth]{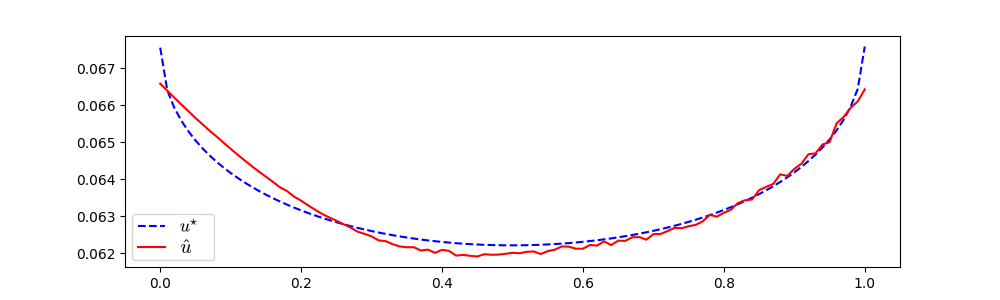}
        \centering \includegraphics[width=\textwidth]{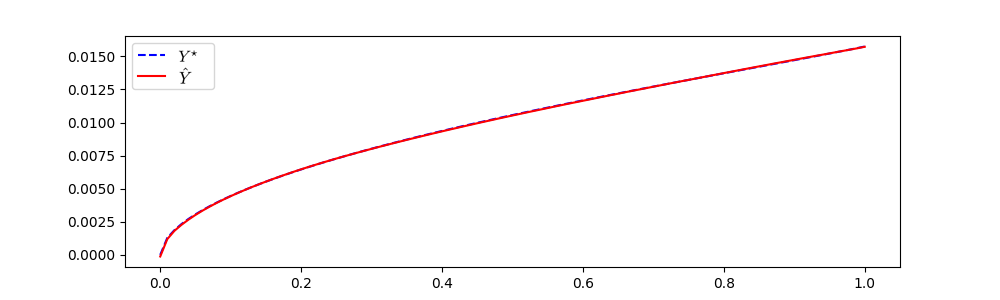}
        \caption{singular power law kernel}
    \end{subfigure}
    \caption{Example trajectories of the OCnet policy $\hat{u}$ with corresponding $\hat{Y}$ (red) compared to the ground truth $u^\star, Y^\star$ (blue) for different propagator kernels with randomly chosen hyperparameters $\theta$ and initial inventory $x$.}
    \label{fig:err_oc_solutions}
\end{figure}

Table~\ref{tab:oc_err} summarizes the relative error of the objective function value in~\eqref{eq:optimal_control_surrogate_discrete} of the trained OCnet policy $\hat{u}, \hat{Y}, \hat{X}$ with respect to the actual value of the objective function of the ground truth optimal policy $u^\star, Y^\star, X^\star$. The three different ICON models (columns) are trained on a single correct dataset (i.e., the prompted in-context examples in the optimization problem are originating from the same propagator kernel type on which the ICON model was pre-trained). We observe that the error values are impressively small. This confirms the effectiveness of our proposed ICON-OCnet approach in finding the optimal execution strategy via few-shot in-context learning with ICON. Sample trajectories of the obtained OCnet selling rate $\hat{u}$ and corresponding price impact $\hat{Y}$ compared to the ground truth optimal execution strategy $u^\star$ and impact process $Y^\star$ for different propagator kernels (with randomly chosen hyperparameters $\theta$ and initial inventory $x$) are shown in Figure~\ref{fig:err_oc_solutions}. 

\begin{table}[ht]
    \centering
    \begin{tabular}{c|c|c|c}
    \hline
    \hline
    & \texttt{ode} & \texttt{ker} & \texttt{sker} \\
    \hline 
    ICON-OCnet & $5.80\times 10^{-8}$ 
     & $6.91\times 10^{-7}$ & $4.55\times 10^{-7}$ \\
    \hline
    \hline
    \end{tabular}
    \caption{Relative error of the optimal control objective function value for ICON models trained on a single correct dataset (rows). The values are averaged over $16$ sets of randomly selected hyperparamters $\theta$ and initial inventories $x$.} 
    \label{tab:oc_err}
\end{table}

\begin{figure}
    \centering
    \begin{subfigure}{0.95\textwidth}
        \centering \includegraphics[width=\textwidth]{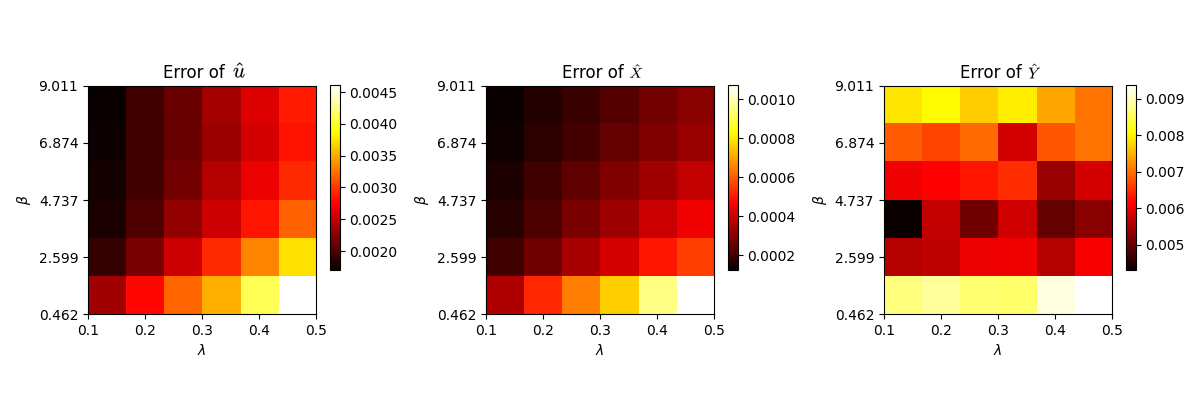}
        \caption{exponential kernel}
    \end{subfigure}
    \begin{subfigure}{0.95\textwidth}
        \centering \includegraphics[width=\textwidth]{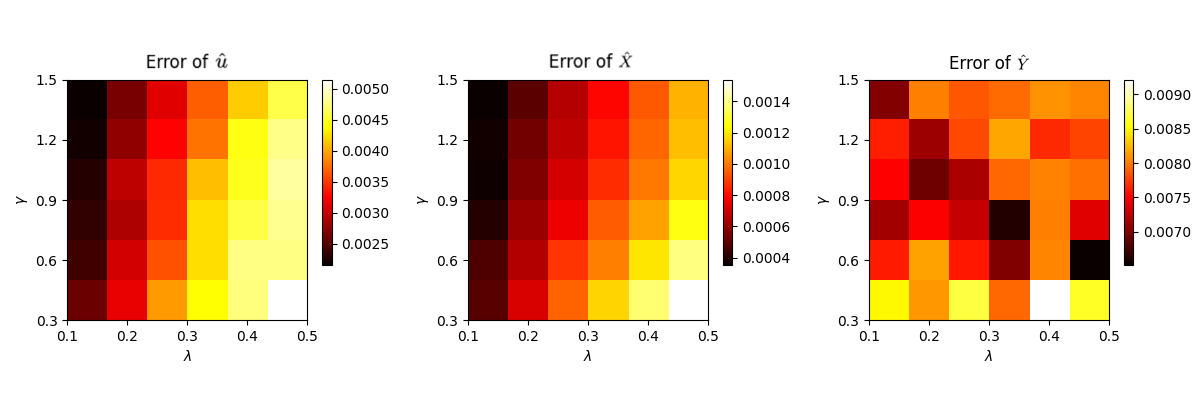}
        \caption{non-singular power law kernel}
    \end{subfigure}
    \begin{subfigure}{0.95\textwidth}
        \centering \includegraphics[width=\textwidth]{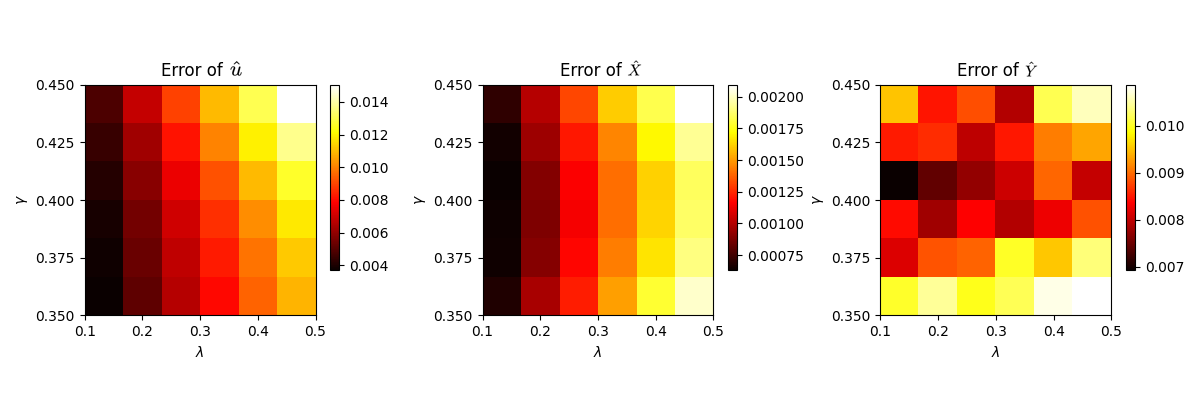}
        \caption{singular power law kernel}
    \end{subfigure}
    \caption{Heatmap for the relative error of the ICON-OCnet policy $\hat{u}$ and corresponding controlled state processes $\hat{X}, \hat{Y}$ with ICON trained on a single dataset (\texttt{ode} top, \texttt{ker} middle, \texttt{sker} bottom). The prompted in-context examples for initializing ICON are in-distribution from \texttt{ode\_t} (top), \texttt{ker\_t} (middle), \texttt{sker\_t} (bottom).}
    \label{fig:heatmap_oc_1}
\end{figure}

\begin{figure}[htbp]
    \centering
    \begin{subfigure}{0.95\textwidth}
        \centering \includegraphics[width=\textwidth]{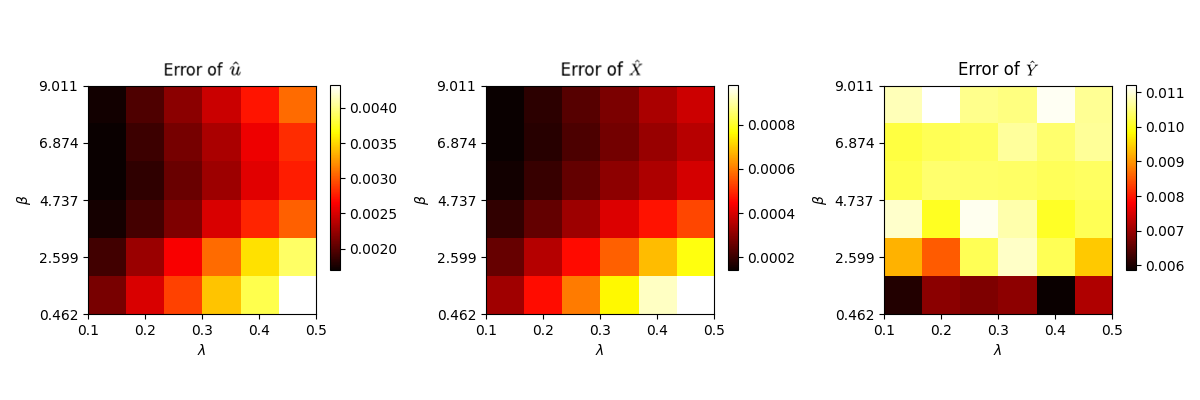}
        \caption{exponential kernel}
    \end{subfigure}
    \\
    \begin{subfigure}{0.95\textwidth}
        \centering \includegraphics[width=\textwidth]{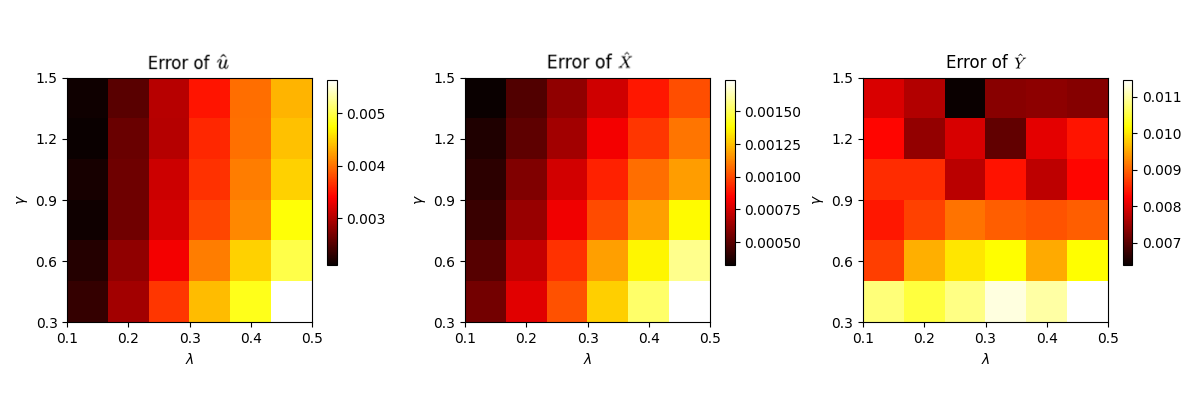}
        \caption{non-singular power law kernel}
    \end{subfigure}
    \\
    \begin{subfigure}{0.95\textwidth}
        \centering \includegraphics[width=\textwidth]{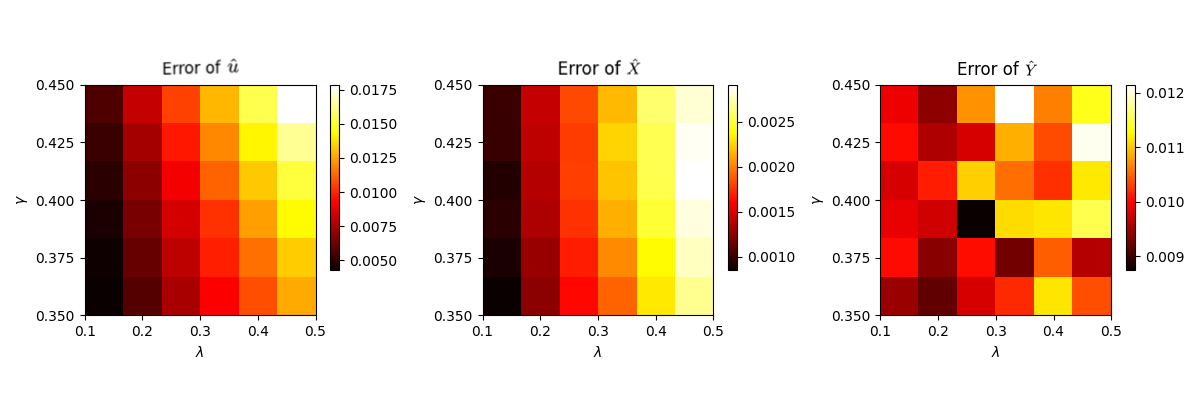}
        \caption{singular power law kernel}
    \end{subfigure}
    \caption{Heatmap for the relative error of the ICON-OCnet policy $\hat{u}$ and corresponding controlled state processes $\hat{X}, \hat{Y}$ with ICON trained on the mixed dataset \texttt{all3}. The prompted in-context examples for initializing ICON are in-distribution from \texttt{ode\_t} (top), \texttt{ker\_t} (middle), \texttt{sker\_t} (bottom).}
    \label{fig:heatmap_oc_2}
\end{figure}

Lastly, we illustrate the influence of the hyperparameters on the error of ICON-OCnet in the heatmaps in Figures~\ref{fig:heatmap_oc_1} and~\ref{fig:heatmap_oc_2}. Similarly to the heatmaps in Section~\ref{subsec:ICONnumerics}, we split the range of each hyperparameter into 6 intervals, and then randomly sample in each box 16 sets of hyperparameters (and initial inventories $x$), for which we train an OCnet policy with the pretrained ICON model (trained on a single dataset in Figure~\ref{fig:heatmap_oc_1} or on the mixed dataset in Figure~\ref{fig:heatmap_oc_2}), prompted with in-distribution examples as context, and compute the relative $\ell^2$ error of $\hat{u},\hat{X},\hat{Y}$ with respect to the corresponding ground truths $u^\star,X^\star,Y^\star$. Overall, the errors are fairly small. As expected, the errors increase slightly when ICON is trained on the mixed dataset \texttt{all3}.

\section{Conclusion and outlook} \label{sec:conclusion}

We introduced ICON-OCnet, a new approach to tackle an optimal stochastic control problem with an unknown operator mapping the control to the controlled state dynamics, which is inferred from examples. To this end, we used In-Context Operator Networks (ICON), a transformer-based neural network architecture introduced in~\cite{YangLiuMengOsher:23}, enhancing data efficiency and model flexibility due to their few-shot and transfer learning capabilities. This paper provides a proof of concept for this general methodology and showcases its efficiency in finding optimal order execution strategies via in-context learning. Specifically, we use trading rates and incurred price impact trajectories as examples to learn the price impact operator (step~1), and then learn the optimal liquidation strategy based on the learned operator (step~2). The price impact environment is unknown and inferred from data prompts (context). We effectively benchmarked our approach against ground-truth solutions for linear propagator models from~\cite{JaberNeuman:25}. ICON-OCnet offers inference with reduced data requirements by merging offline pre-training with online few-shot learning, finding the model that is closest to the prompted examples and eliminating the need for retraining when new contexts arise.

Our work paves the way for future research in several directions: First, it is very sensible to test the performance of ICON-OCnet on a larger class of price impact models, including non-linear models studied in~\cite{AbiJaberEtAl:25}. This requires a more involved benchmarking procedure since ground-truth optimal execution strategies are no longer explicitly available but can only be computed approximately via numerical iterative schemes. Moreover, it is very interesting to investigate the training of ICON based on noisy in-context time series sample data pairs $(u,P)$ as described in Section~\ref{subsec:training:uP:based}, including non-martingale dynamics for the unaffected price process, rather than on exact trajectory pairs $(u,Y)$. This will require a much larger training set and more in-context examples to filter out the price impact. It might also require some degree of regularization to prevent overfitting. In a similar vein, we performed our offline training and online few-shot prompting using synthetic data. Of course, it would be interesting to study the training of ICON on real trade execution data. More precisely, due to the potential scarcity of the latter, it is conceivable to study the offline training on synthetic model-based data augmented by historical trade execution data, as well as using real trade execution data for the online in-context few-shot inference. However, this requires proprietary trading data on meta-order execution, which is notoriously hard to come by. Last but not least, it is also very appealing to study our sample-based in-context learning methodology for general path-dependent stochastic optimal control problems with general noise-driven state dynamics.

\section*{Acknowledgement}
We would like to express our gratitude to Liu Yang for valuable discussions. T.M.~and S.O.~are supported by ONR MURI N00014-20-1-2787.

\bibliographystyle{alpha}
\bibliography{literature}

\end{document}